\documentclass[twocolumn,superscriptaddress,pra,aps,showpacs,floatfix]{revtex4-1}
\usepackage{amsfonts}
\usepackage{amsmath}
\usepackage{amssymb}
\usepackage{graphicx}

\newcommand{\ket}[1]{|#1\rangle}
\newcommand{\bra}[1]{\langle #1|}
\newcommand{\order}[1]{\mathcal O(#1)}
\newcommand{\braket}[2]{\langle #1| #2 \rangle}


\begin{document}

\title{Dynamical crystal creation with polar molecules or Rydberg atoms in optical lattices}

\author{J~Schachenmayer} 
\affiliation{Institute for Theoretical Physics, University of Innsbruck, A-6020
Innsbruck, Austria}
\affiliation{Institute for Quantum Optics and Quantum Information of the Austrian Academy
of Sciences, A-6020 Innsbruck, Austria}
\author{ I~Lesanovsky} 
\affiliation{School of Physics and Astronomy, The University of Nottingham,
University Park, Nottingham NG7 2RD, UK}
\author{A~Micheli}
\affiliation{Institute for Theoretical Physics, University of Innsbruck, A-6020
Innsbruck, Austria}
\affiliation{Institute for Quantum Optics and Quantum Information of the Austrian Academy
of Sciences, A-6020 Innsbruck, Austria}
\author{A~J~Daley}
\affiliation{Institute for Theoretical Physics, University of Innsbruck, A-6020
Innsbruck, Austria}
\affiliation{Institute for Quantum Optics and Quantum Information of the Austrian Academy
of Sciences, A-6020 Innsbruck, Austria}

\date{\today}

\begin{abstract}
  We investigate the dynamical formation of crystalline states with
  systems of polar molecules or Rydberg atoms loaded into a deep
  optical lattice. External fields in these systems can be used to
  couple the atoms or molecules between two internal states: one that
  is weakly interacting and one that exhibits a strong dipole-dipole
  interaction. By appropriate time variation of the external fields,
  we show that it is possible to produce crystalline states of the
  strongly interacting states with high filling fractions chosen via
  the parameters of the coupling. We study the coherent dynamics of
  this process in one dimension (1D) using a modified form of the
  time-evolving block decimation (TEBD) algorithm, and obtain
  crystalline states for system sizes and parameters corresponding to
  realistic experimental configurations. For polar molecules these
  crystalline states will be long-lived, assisting in a
  characterization of the state via the measurement of correlation
  functions. We also show that as the coupling strength increases in
  the model, the crystalline order is broken. This is characterized in
  1D by a change in density-density correlation functions, which decay
  to a constant in the crystalline regime, but show different regions
  of exponential and algebraic decay for larger coupling strengths.
\end{abstract}

\maketitle

\section{Introduction and overview}

Recent experimental progress in controlled excitation of Rydberg atoms
from cold gases \cite{re1,re2,re3,re4,re5,re6} and in the production
of ultracold polar molecules \cite{Ni08} has opened new opportunities
for the study of systems with dipole-dipole interactions. Key features
of these systems include the ability to control these interactions by
varying the external fields, as well as the strong dependence of the
interaction strength on the internal state of the atoms or
molecule. An important example of this dependence is the Rydberg
blockade mechanism \cite{Jaksch00,Lukin01} in which an atom in an
excited Rydberg state can prevent nearby atoms from being excited to
Rydberg states, because the strong interaction between the two excited
atoms makes subsequent excitations non-resonant. This mechanism has
been recently demonstrated in experiments \cite{Urban09,Gaetan09}, and
has important potential applications to the production of fast quantum
gates with two or several neutral atoms
\cite{Jaksch00,Lukin01,Mueller09,Saffmann09}. Moreover, the blockade
produces rich and intriguing collective dynamics when atoms in a dense
gas are excited to Rydberg states \cite{re1, re2, re3, re4, re5, re6,
  Sun08, Olmos08, Olmos10}. This cooperative behaviour is manifest in
a suppression of the proportion of excited atoms with increasing
density (due to the presence of more atoms within the blockade
radius), and also in an enhanced laser coupling to collective Rydberg
excitations.

In the context of Rydberg atoms, there has been a lot of interest in
the possible production of crystalline states, where a small fraction
of atoms in excited Rydberg states arrange themselves in a regular
array \cite{Weimer08, Weimer09, Olmos09, Pohl10}. In particular, when
ground and excited states are coupled by a laser field, the resulting
effective models correspond to crystalline states in the limit of weak
coupling, with the fraction of excited atoms determined by a
competition between the interaction energy and the detuning of the
laser field from the atomic transition.

Motivated by this experimental and theoretical pro\-gress, we consider
here a collection of either polar molecules or neutral atoms loaded
into a deep optical lattice, e.g., with one particle per site. We
assume that tunnelling in the lattice is negligible on the timescale
of the experiment. Using dressed rotational states for the polar
molecules or Rydberg states for the atoms we show how a microscopic
model can be realized where two internal states of the particles are
accessed, one of which exhibits a large dipole-dipole interaction,
whilst the other is weakly interacting. We show that this system
allows the realisation of crystalline formations in the internal
states, with a periodicity related to the lattice period by a rational
fraction. The regularity of the system makes it possible to realize
crystals with high filling factors of strongly interacting states. In
addition, the use of polar molecules makes it possible for the
crystalline states to be long-lived, aiding in the measurement of
characteristic correlation functions. We show that in a 1D tube,
increasing the coupling strength between the internal states at a
fixed detuning will break the crystalline order. This is characterized
in 1D by a change in the density-density correlation (DDC) functions,
which decay to a constant in the crystalline regime, but show
different regions of exponential and algebraic decay for larger
coupling strengths.

We study in detail the dynamical preparation of states with
crystalline order of different periodicities in the internal states by
an appropriate time-variation of microwave or laser couplings. As was
shown in previous studies for Rydberg atoms in an optical lattice
modelled by nearest-neighbour interactions \cite{Sun08, Olmos08,
  Olmos10}, it is not possible to produce crystalline states simply by
switching on the coupling field. Instead, this is made possible using
an adiabatic variation of the parameters in the effective model. Using
an extended version of the Time-Evolving Block Decimation (TEBD)
algorithm \cite{Vidal03, Vidal04}, we calculate the coherent
time-dependent many-body dynamics of the excitations. We show that for
system sizes and parameters corresponding to realistic experimental
configurations with either polar molecules or Rydberg atoms we obtain
fidelities $>99\%$ compared with the ideal target states, i.e., the
ground states of the effective model. Whilst the fidelity could be
reduced by incoherent processes, e.g. decay of excited states, we
estimate that these rates are small, and that the basic structure of
the crystalline state will be robust. Note that the time-dependent
preparation of crystalline phases here follows the same philosophy as
that for production of Rydberg crystals in reference~\cite{Pohl10}. In
focussing on the case of atoms or molecules initially trapped in a
lattice, we concentrate here on the case with a high proportion of
particles in the strongly interacting case. This is in contrast to
reference~\cite{Pohl10}, which treats the case of a dilute gas of
Rydberg excitations that is more relevant for a frozen Rydberg gas not
trapped in a lattice. Our study of the correlation functions in the
crystalline and non-crystalline phases is, however, also relevant for
the case of \cite{Pohl10}.

The remainder of the paper is organized as follows: in
section~\ref{sec:real-long-range}, we introduce the effective
long-range spin-model that describes both the Rydberg atoms and polar
molecules in appropriate parameter regimes. Then we briefly discuss in
section~\ref{sec:numerical-method} extensions to the TEBD algorithm
that we use to study the model. In
section~\ref{sec:ground-states-system}, we analyse the ground states
of this effective model and identify a target state of our
time-dependent excitation process. The ground states are characterized
as belonging to different quantum phases via the characteristic
behaviour of DDC functions. In
section~\ref{sec:dynam-cryst-state}, we discuss a scheme to prepare
crystalline states with high fidelity by an appropriate
time-dependence of the external coupling field parameters.  In
section~\ref{sec:phys-impl-long}, details of two different physical
realizations of the effective model, with polar molecules
(section~\ref{sec:polar-molecules-an}) and Rydberg atoms
(section~\ref{sec:rydb-excit-an}) are provided. Finally,
section~\ref{txt:outlook}, provides the conclusion and outlook.

\section{The System}
\label{sec:real-long-range}

We consider particles loaded into an optical lattice in a regime where
the lattice is sufficiently deep that tunnelling can be neglected on
the timescale of the experiment. We assume that the particles have two
internal states, one of which exhibits a strong dipole-dipole
interaction, whilst the other is weakly interacting. This can be
realized for realistic experimental parameters either
with \begin{enumerate}
\item{\emph{Polar molecules} in an external electric field, where a
    microwave coupling of rotational states can give rize to dressed
    states with different dipole-dipole interactions, or}
\item{\emph{Neutral atoms} in an external electric field coupled via a
    laser to excited Rydberg states, which in contrast to the ground
    states exhibit a strong dipole-dipole interaction.}
\end{enumerate}
Below we summarize the key points of each of these
implementations. Further details for polar molecules and neutral atoms
coupled to excited Rydberg states are given in
section~\ref{sec:polar-molecules-an} and
section~\ref{sec:rydb-excit-an} respectively.

\subsection{Polar molecules}

Polar molecules in their electronic-vibrational ground-state manifold
possess a number of features that make them appealing to design and
implement specific spin-models. In addition to their electric
structure, allowing e.g. for optical trapping, they have a rich
internal (rotational) symmetry, with level spacings in the GHz regime,
and negligible decoherence. They possess a permanent dipole-moment
$d_0$ (typically on the order of a few Debye \cite{Krems09, Carr09})
along the internuclear axis, which makes it possible to strongly
couple their rotational excitations via static and microwave fields
and gives rise to dipole-dipole interactions between different
molecules.

\begin{figure}[htb]
  \centering
  \includegraphics[width=0.45\textwidth]{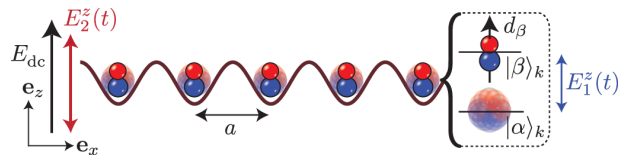}
  \caption{Schematic illustration of a system of polar molecules
    loaded into subsequent sites of an 1D optical lattice with site
    separation $a$. The molecules are aligned by a static electric
    field in $\mathbf{e}_z$-direction, $E_\mathrm{dc}$, and exposed to
    two time-dependent microwave fields $E^z_1(t), E^z_2(t)$, both
    linearly polarized along the $z$-axis. } \label{fig:pmsetup_1}
\end{figure}

We consider a 1D optical lattice (lattice spacing $a$) along the
$\mathbf{e}_x$ axis with one polar molecule trapped at each lattice
site in the lowest vibrational state $\phi_k(\mathbf{x})$ (see
figure~\ref{fig:pmsetup_1}), which can be realized either by loading
an insulator state of polar molecules, or beginning deep in a Mott
Insulator state with two atoms per site and associating them to form
polar molecules \cite{Jaksch02}.  Without external fields, the
rotational states of the molecules do not possess a net dipole
moment. However, finite dipole moments along the $z$-axis in the
rotational states can be induced by applying a static external
electric field $E_\mathrm{dc}$ in that direction. The key idea of our
implementation is that a suitable choice of $E_2^z(t)$ makes it
possible to cancel the induced dipole moment in one of the dressed
states, and we denote the complete state of an atom in this dressed
state at site $k$ as $\ket{\alpha}_k$. The single molecule ground
state, which obtains a strong dipole moment from the static electric
field, is denoted as $\ket{\beta}_k$. In the case that $E_\mathrm{dc}$
and $E_2^z(t)$ are properly tuned, two molecules at distinct sites $k$
and $l$ that are in the states $\ket{\beta}_k$ and $\ket{\beta}_l$
lead to an energy shift due to the dipole-dipole interaction potential
of $V_\mathrm{dd}/|k-l|^3$. In section~\ref{sec:polar-molecules-an} we
give example experimental parameters showing that an energy shift
$V_\mathrm{dd}\approx 0.7 d_0^2/4 \pi \epsilon_0 a^3 \approx 10 \hbar
\, \mathrm{kHz} $ can be achieved.  Note that we can therefore assume
that the molecules remain in their motional ground state, since the
band separation (trapping frequency) $\omega_T$ is typically of the
order of many tens of kHz up to $100$ kHz in a realistic scenario.
Finally, the states $\ket{\beta}_k$ and $\ket{\alpha}_k$ on all sites
$k$ are coupled by a weak microwave field $E_1^z(t)$ with Rabi
frequency $\Omega$ and detuning $\delta$, which gives rise to the
effective microscopic many-body model that is described in
section~\ref{sec:effmod} below.

\subsection{Neutral atoms coupled to Rydberg states}

Neutral atoms in an external electric field exhibit strong
dipole-dipole interactions when in certain Rydberg states with
relatively large principal quantum number. For alkali atoms in states
with principle quantum number $n=17$, it is possible to obtain a
dipole moment of more than one kiloDebye (see
section~\ref{sec:rydb-excit-an} for details). By finding isolated Rydberg
states, the internal dynamics of each atom can be reduced to a two
level system, where weakly interacting atoms in the ground state are
coupled via a laser to these states with large dipole-dipole
interactions.

We assume that either atoms are loaded into a deep optical lattice
with tight radial confinement, so that we have one atom per site along
a 1D tube, or that we have weak radial confinement and many atoms per
site (as occurs, e.g., when an optical lattice is applied along only
one direction). In each case, we can assume that the ground-state
atoms are confined to a single modefunction at each given lattice
site, $\phi_k(\mathbf{x})$. In the case of a single atom per site,
this is the Wannier function for the lowest Bloch band, which is
analogous to the case of the polar molecules described above. For many
atoms per site, interactions can populate higher bands, and
$\phi_k(\mathbf{x})$ will depend on the interactions. However, we
assume that the timescale for excitation to the Rydberg state is much
shorter than timescales corresponding to the atomic motion, and that
therefore we can neglect population of any modefunction other than the
initial modefunction $\phi_k(\mathbf{x})$ for either the ground or
Rydberg states.  We then assume that we can have at most one
excitation per lattice site, which in the case of large single-site
populations is made possible by the Rydberg blockade mechanism, when
we ensure that the Blockade radius is larger than the occupied region
at any single lattice site (see section~\ref{sec:rydb-excit-an}). This
makes it possible to describe the system with an effective spin-1/2
model, assigning a state on each lattice site to the existence or
non-existence of an atom in an excited Rydberg state on that site,
labelled for site $k$ by the states $\ket{\beta}_k$ and
$\ket{\alpha}_k$, respectively.

In comparison with polar molecules, the dipole-dipole interaction of
the strongly interacting state for Rydberg atoms is much larger. If we
choose a principal quantum number $n=14$, then $V_{\mathrm{dd}}\approx
7 \hbar \, \mathrm{GHz}$ (assuming a lattice spacing of
$a=400\,\mbox{nm}$, see section~\ref{sec:rydb-excit-an} for more
details). This corresponds to much faster experimental
timescales. However, the polar molecules are much longer lived than
atoms in excited Rydberg states, and open up different opportunities
for experimental observation of the crystalline states we discuss
here.

\subsection{The effective model}
\label{sec:effmod}

In either the polar molecule or neutral atom case we can derive the
same effective microscopic model for the system in a frame rotating
with the frequency of the field coupling the two internal states.  We
assume that at most one particle per site can be in the strongly
interacting state, and represent the two possible states (weakly and
strongly interacting) at each lattice site as two spin states. We
label these states for site $k$ as $\{ \ket{\beta}_k, \ket{\alpha}_k
\}$, where $\ket{\beta}_k$ is the state exhibiting strong
dipole-dipole interactions, and $\ket{\alpha}_k$ the non-interacting
state. The Hamiltonian can then be formulated as ($\hbar \equiv 1$)
\begin{align} \label{eq:hamiltonian}
  \hat H
  =
  \Omega \sum_k  \hat \sigma_k^x
  -
  \delta \sum_k \hat n_k
  +
 \frac{V_{\mathrm{dd}}} {2} \sum_{k \neq m} \frac{\hat n_k \hat n_m} {|k-m|^3}
 ,
\end{align}
with $\hat n_k\equiv \ket{\beta}\bra{\beta}_k$ being the local
``number operator'' for the interacting state at site $k$. The Pauli
matrices are defined via $\hat
\sigma_k^x\equiv\ket{\beta}\bra{\alpha}_k + \ket{\alpha}\bra{\beta}_k$
and $\hat \sigma_k^z\equiv\ket{\beta}\bra{\beta}_k -
\ket{\alpha}\bra{\alpha}_k$, where the latter is related to the number
operators via $\hat n_k=(\hat \sigma_k^z +1)/2$. $\Omega$ denotes the
effective Rabi frequency and $\delta$ the detuning of the laser, which
excites the particles. $V_{\mathrm{dd}}$ is the dipole-dipole
interaction energy shift between neighbouring sites. Note that we
assume the particles to remain in the motional states
$\phi_k(\mathbf{x})$, and also neglect corrections to the $|k-m|^{-3}$
decay due to the finite width of $\phi_k(\mathbf{x})$ (see
section~\ref{sec:phys-impl-long} for more details).

Below we will first study the ground state properties of this model,
which will give us target states for the dynamical preparation
process. We will then study the time dependence of this system
starting from a many-body state where all of the particles are weakly
interacting ($\bigotimes_k \ket{\alpha}_k$), and investigate the
adiabatic preparation of crystalline states as the parameters are
varied.

It is interesting to note that for an infinite system, the Hamiltonian
can be written as
\begin{align} \label{eq:spinhamiltonian}
  \hat H_S
  =
  \Omega \sum_k  \hat \sigma_k^x
  &+
  \frac{1}{2}(\zeta(3) V_{\mathrm{dd}} - \delta) \sum_k \hat \sigma^z_k \nonumber \\
  &\qquad
  +
 \frac{V_{\mathrm{dd}}} {8} \sum_{k \neq m} \frac{\hat \sigma^z_k \hat \sigma^z_m} {|k-m|^3} 
.
\end{align}
where we have removed constant terms, and $\zeta(3) \equiv \sum_k
k^{-3}\approx 1.202$ denotes Ap\'{e}ry's constant
\cite{Abramowitz}. Note that in the case of
$\delta=\zeta(3)V_\mathrm{dd}$, the system is equivalent to an Ising
model in a transverse field with dipolar interactions. Ground states
of this model have been also studied in \cite{Deng05}.

\section{Numerical Method}
\label{sec:numerical-method}

To calculate the many-body dynamics and ground states of
Hamiltonian~(\ref{eq:hamiltonian}), we make use of the TEBD algorithm
as introduced in \cite{Vidal03, Vidal04}. This algorithm makes
possible the near-exact integration of the Schr\"odinger equation for
1D lattice and spin Hamiltonians based on a matrix product state
ansatz, and has been applied to a range of such Hamiltonians with
next-neighbour interactions. There are also applications of matrix
product state algorithms to dissipative systems
\cite{Verstraete04,Zwolak04,Daley09}, and these methods have been
incorporated within density matrix renormalization group algorithms
\cite{Daley04, White04}. There is also a strong effort to extend these
ideas to higher dimensions \cite{MPSAdvPhys}.

The modification in our work is the extension of these methods to the
treatments of finite-range interactions. For the purposes of this
study, we include long-range interactions up to an arbitrary distance
of $l$ sites, by implementing a routine for swapping site indices in
the matrix product state representation. This adds a computational
cost of $\order{l}$ basic operations compared to the standard TEBD
calculation. Here we work with finite systems of size $N$, and always
properly represent interactions over the full length of the system,
i.e. $l=N-1$. This method of swapping also allows us to consider
periodic boundary conditions for the dipole-dipole interactions. Note
that we always perform convergence tests in the matrix product state
bond-dimension $\chi$ and other numerical parameters
\cite{footnote1}. For larger system sizes, swapping sites can become
inefficient, in part due to the higher required values of $\chi$. To
implement interaction terms over substantially larger distances than
in this study, more efficient algorithms should be used, e.g.,
algorithms involving use of matrix product operators
\cite{Mcculloch07, Crosswhite08, Pirvu10}.

\section{Ground states of the effective model}
\label{sec:ground-states-system}

We begin by studying the key ground state properties of the effective
model of Hamiltonian~(\ref{eq:hamiltonian}). This will allow to us to
determine in which parameter regimes crystalline order will appear,
and to characterize the states that will be the target states of the
dynamical preparation process discussed next in
section~\ref{sec:dynam-cryst-state}.

\begin{figure}[htb]
\centering
\includegraphics[width=0.45\textwidth]{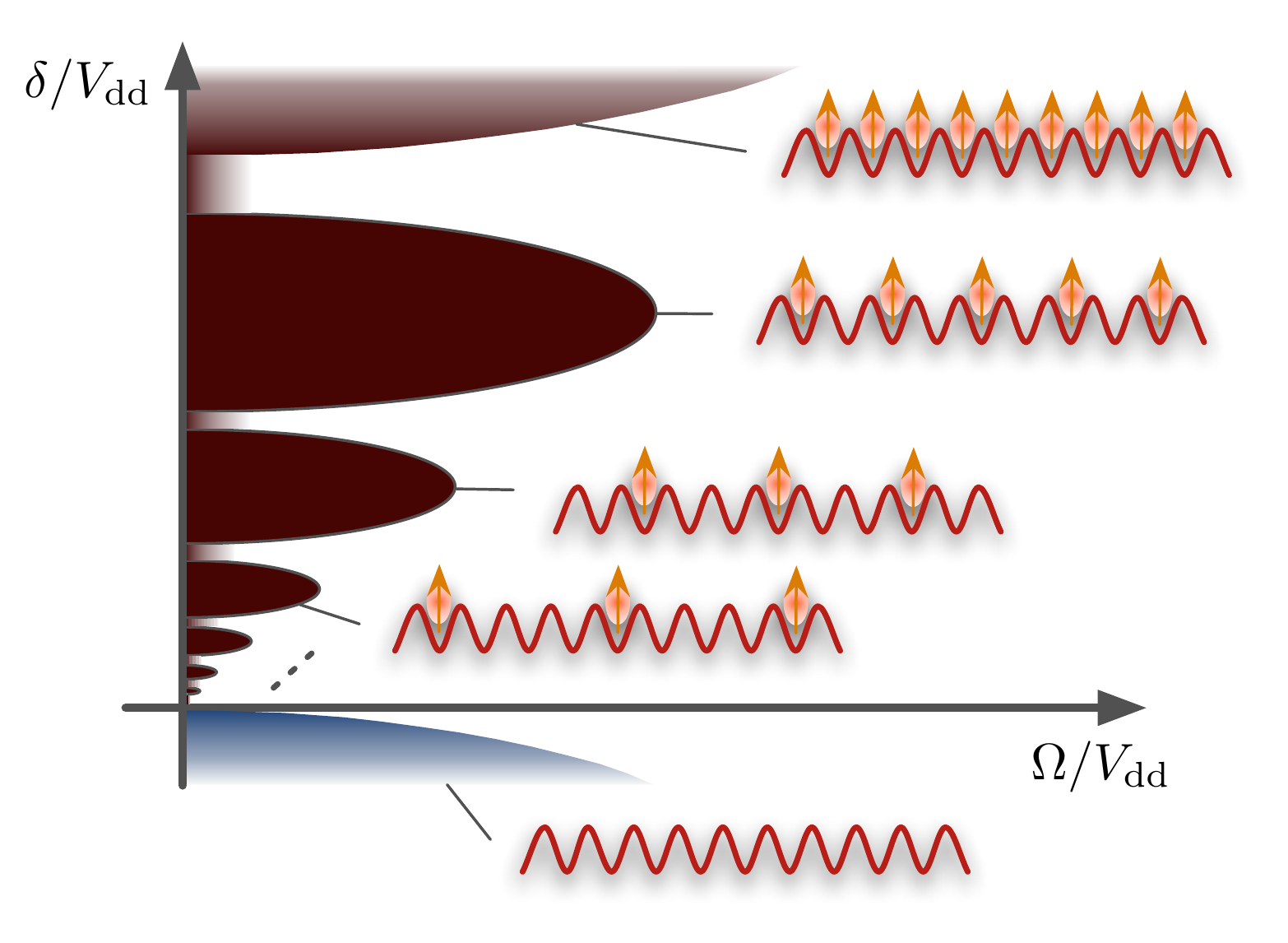}
\caption{A sketch of the key features of the phase diagram for the
  system with Hamiltonian (\ref{eq:hamiltonian}). For $\delta<0$ and
  $\Omega/V_\mathrm{dd} \ll 1$, the ground state involves all of the
  atoms in the weakly interacting state in order to minimize the
  interaction energy (marked by the blue lobe). As the detuning
  $\delta>0$ is increased, it becomes favourable to add excitations to
  the system, but this competes with an increasing interaction
  energy. This competition gives rise to crystalline phases with
  periodicities determined by this competition (marked by the brown
  lobes). For large $\Omega/V_\mathrm{dd}$, the crystalline order is broken,
  but some particles exist in excited states (white area).}
\label{fig:phase_diag_sketch}
\end{figure}

The Hamiltonian we study here has strong similarities to another model
treated in several recent studies, in which the dipole-dipole
interactions were replaced with van der Waals interactions arising for
certain Rydberg systems \cite{Weimer08,Weimer09, Olmos09, Olmos10,
  Pohl10}. In \cite{Weimer08} it was shown that for $\Omega=0$ that
system reduces to a classical spin model exhibiting a second order
phase transition from a paramagnetic to a crystalline phase at
$\delta=0$, and we observe similar behaviour here. Due to the dipolar
long range character we expect crystalline states with different
periodicities of the excitations to the strongly interacting state to
appear for $\delta>0$ and small values of $\Omega$. The detuning plays
the role of a chemical potential for strongly interacting states (or
external magnetic field in the spin model), and it becomes favourable
to excite states with a dipole-dipole interaction on the
lattice. However, the long-range interactions will compete
energetically with the detuning, leading to crystal periodicities that
decrease with increasing $\delta/V_{ \rm dd}$. Below we show that as
the coupling $\Omega/V_{ \rm dd}$ is increased, the crystalline order
is broken, in favour of a paramagnetic phase in the spin language,
with no long range density-density ordering. These key features of the
phase diagram are sketched in figure~\ref{fig:phase_diag_sketch}.

\begin{figure}[htb]
\centering
\includegraphics[width=0.45\textwidth]{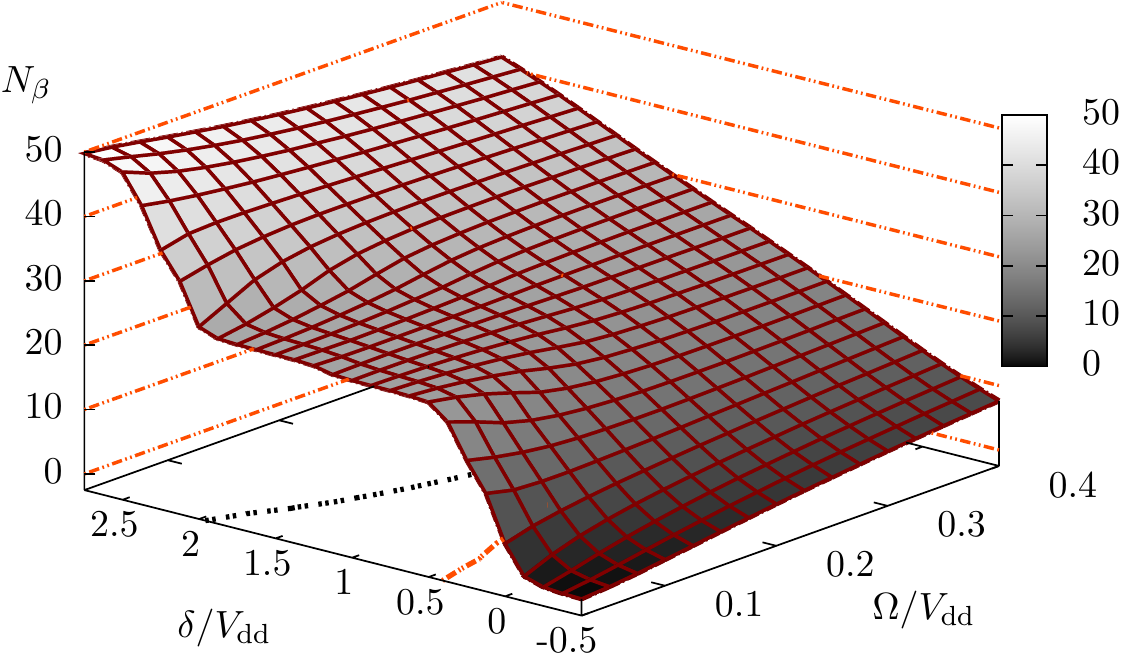}
\caption{Occupation number of strongly interacting states, $N_\beta$,
  for ground states of the effective model in a $50$-site system.  The
  Hamiltonian is given through equation~(\ref{eq:hamiltonian}) and open
  boundary conditions are assumed. Results are shown for several
  values of detuning $\delta$ and Rabi frequency $\Omega$, and
  plateaus of zero, half, and full filling of the lattice are clearly
  visible for $\Omega \lesssim 0.1 V_{\mathrm{dd}}$ (see text). The
  contours of $N_\beta=23$ and $N_\beta=27$ are drawn as lines on the
  $\delta$-$\Omega$ plane.}
\label{fig:phase_diag}
\end{figure}

We compute the ground state of equation~(\ref{eq:hamiltonian}) using the
TEBD algorithm in several different parameter regimes. As a first
step, we look at the total number of particles in a strongly
interacting state $\ket{\beta}$,
\begin{align}
N_\beta\equiv \sum_k^N \langle \hat n_k \rangle.
\end{align}
In figure~\ref{fig:phase_diag} we present a plot of $N_\beta$ for
several detunings $\delta$ and effective Rabi frequencies $\Omega$. We
plot $N_\beta$ on a grid with spacings $\Delta \delta = 0.125
V_{\mathrm{dd}}$ and $\Delta \Omega = 0.025 V_{\mathrm{dd}}$.

As expected, for small $\Omega/V_\mathrm{dd}$ there is essentially no
occupation of the strongly interacting states in regions with negative
detuning, because a state with no $\ket{\beta}_k$ excitations is
energetically favoured.  In contrast, a positive detuning leads to a
reduction of the total energy when occupations of strongly interacting
states are added to the system, and in regions with large positive
detuning we find states with occupations of $\ket{\beta}_k$ on all
lattice sites. In the region between we observe varying excitation
numbers depending on the choice of parameters. In particular, there is
a large plateau in the region $0.5 V_{\mathrm{dd}} \lesssim \delta
\lesssim 2 V_{\mathrm{dd}}$ and $\Omega \lesssim 0.1 V_{\mathrm{dd}}$,
in which $N_\beta$ remains close to $N/2=25$.  In the plot the
contours of fillings slightly smaller ($N_\beta=23$) and larger
($N_\beta=27$) than $N/2$ are drawn as lines on the $\delta$-$\Omega$
plane. Note that the $N/2$ plateau lobe is centred around a value
slightly larger than $\delta=1$, which is approximately the point
where the second term in equation~(\ref{eq:spinhamiltonian}) vanishes and
our system becomes an effective Ising spin model.

The half filling plateau disappears with increasing $\Omega$ and we
find a linear interpolation between regions with zero and full filling
in this regime. This is consistent with a breakdown of the crystalline structure,
and we note that in the limit $\Omega \to \infty$ for finite
$\delta$ we expect the filling to be constant at $N/2$ since the state
$\bigotimes_k^N \left ( \ket{\beta}_k - \ket{\alpha}_k \right) / \sqrt{2}$
becomes the ground state of the system.

Note that the grid is too coarse-grained to resolve plateaus of
smaller fractions of the filling. However, we observe several kinks in
the particle number at small values of $\Omega/V_\mathrm{dd}$. We show
below, e.g., that the feature at $\delta\approx 0.25 V_{\mathrm{dd}}$
corresponds to a crystalline state with filling $N/3$. This will
verify the lobe-like structure as depicted in
figure~\ref{fig:phase_diag_sketch}, however we find that the plateau
for filling $N/3$ is significantly smaller, at values of
$\Omega\approx 10^{-2} V_\mathrm{dd}$. For values of $\delta \sim 2.25V_\mathrm{dd}$
we also observe some kinks, which correspond to regular patterns in
the lattice with two subsequent sites being in the strongly
interacting state, followed by one site being in the state $\ket{\alpha}_k$.

\begin{figure}[htb]
  \centering
  \includegraphics[width=0.45\textwidth]{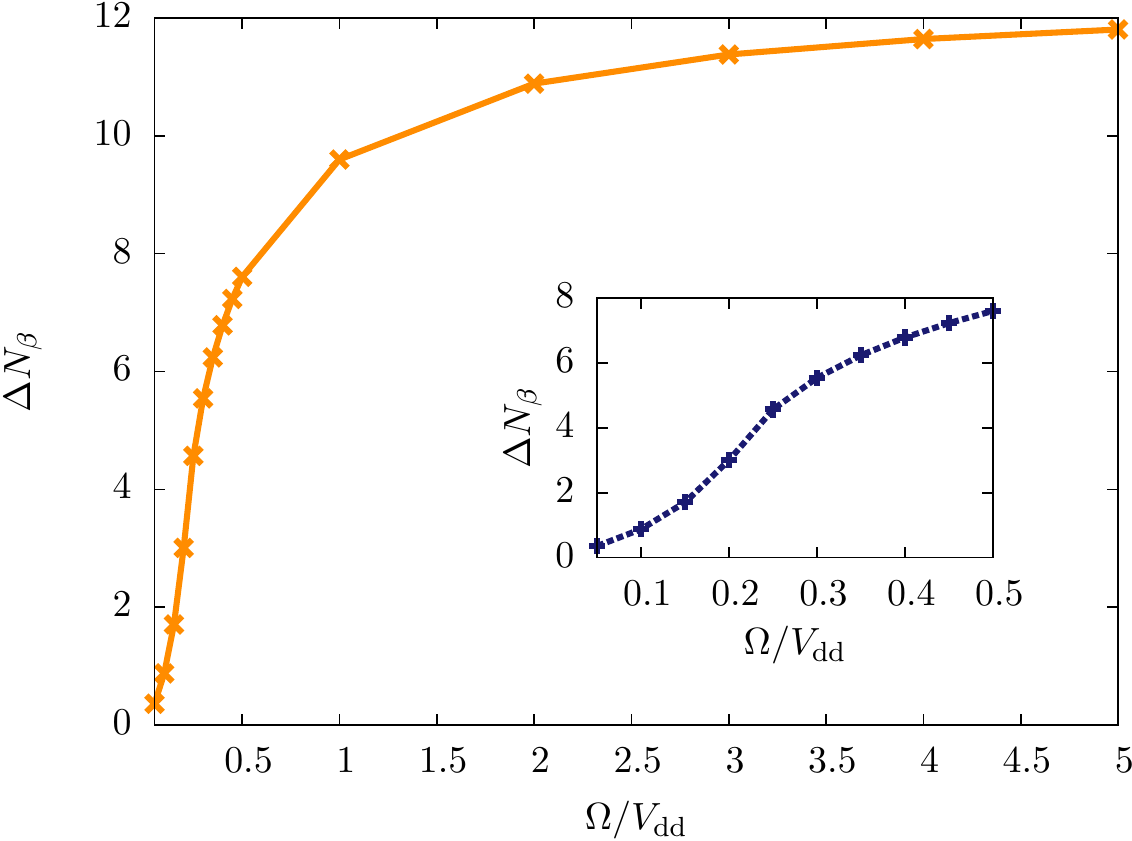}
 \caption{Fluctuation of the total occupation number of the strongly
    interacting state in a $50$ site system with open boundary
    conditions as a function of $\Omega$ for $\delta=1.1
    V_{\mathrm{dd}}$. The inset shows a zoom into the region $0.05
    V_{\mathrm{dd}} \leq \Omega \leq 0.5
    V_{\mathrm{dd}}$.}\label{fig:totnumfluc}
\end{figure}

In order to illustrate the claim that the plateaux seem to belong to
crystalline states, we can investigate the fluctuation in the total
occupation number of the strongly interacting state,
\begin{align}
  \label{eq:numberfluctuation}
  \Delta N_\beta \equiv
  \sum_{k,l}^N \langle \hat n_k \hat n_l \rangle-
  N_\beta^2.
\end{align}
In figure~\ref{fig:totnumfluc} we show $\Delta N_\beta$ for $\delta=1.1
V_{\mathrm{dd}}$ and $0.05 V_{\mathrm{dd}} \leq \Omega \leq 5
V_{\mathrm{dd}}$. We find that $\Delta N_\beta$ decreases
significantly with decreasing $\Omega$ and a crossover behaviour is
centred around $\Omega \approx 0.2 V_{\mathrm{dd}}$. For $\Omega
\lesssim 0.2$ a crystalline state with nearly no fluctuations $\Delta
N_\beta \lesssim 3 \ll N = 50$ is present. For large $\Omega$, $\Delta
N_\beta$ converges to a value close to $N/4=12.5$, which is the
correct result for finite $\delta$ and $\Omega \to \infty$, i.e. where
$\bigotimes_k^N \left ( \ket{\beta}_k - \ket{\alpha}_k \right) /
\sqrt{2}$ becomes the exact ground state of the system. Note that the
quantity $\Delta N_\beta$ is related to the Mandel Q-factor $Q \equiv
\Delta N_\beta / N_\beta - 1$, which has been measured in recent
experiments for the number of Rydberg excitations in small ensembles
of cold atoms \cite{Cubel05} and quantifies the deviation from a
Poissonian number statistic \cite{Ates06}. 

To study ground states in more detail for different parameter regimes,
and demonstrate clearly the crystalline behaviour we evaluate DDC
functions, defined as
\begin{align} \label{eqn:DDC} 
  \mathcal{G}^{[i]}_k
  \equiv
  \langle \hat n_i \hat n_{i+k} \rangle 
  - \langle \hat n_i \rangle\langle \hat n_{i+k} \rangle.
\end{align}
In a finite system with open boundary conditions, the exact values of
$\mathcal{G}^{[i]}_k$ will differ from the site $i$ where the DDC is
evaluated. However, the decay behaviour of the correlation functions
for large $k$ should be independent of the site index $i$ in a
sufficiently large system. To distinguish ground state phases by their
decay behaviour, we will first calculate ground states of a system with
periodic boundary conditions, in which the DDC becomes site
independent. We then drop the site index and write $\mathcal{G}_k$.

\begin{figure}[tb]
  \centering
  \includegraphics[width=0.45\textwidth]{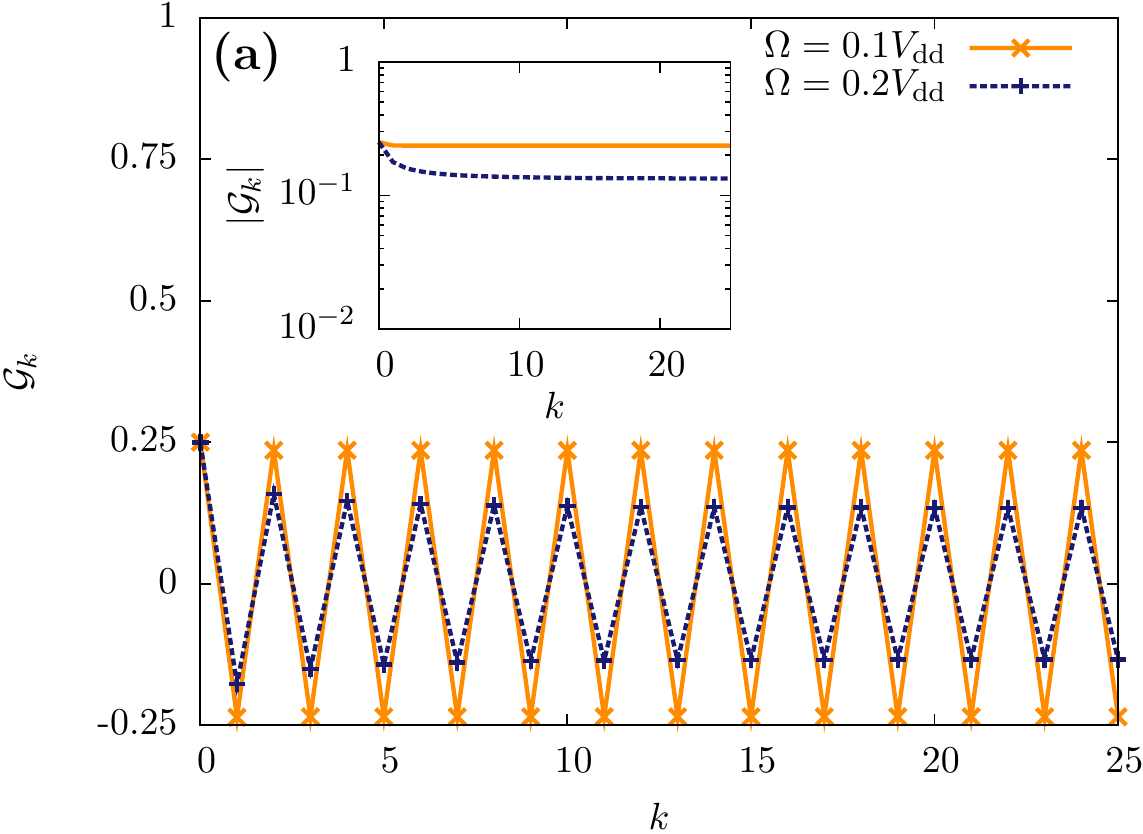}
  \includegraphics[width=0.45\textwidth]{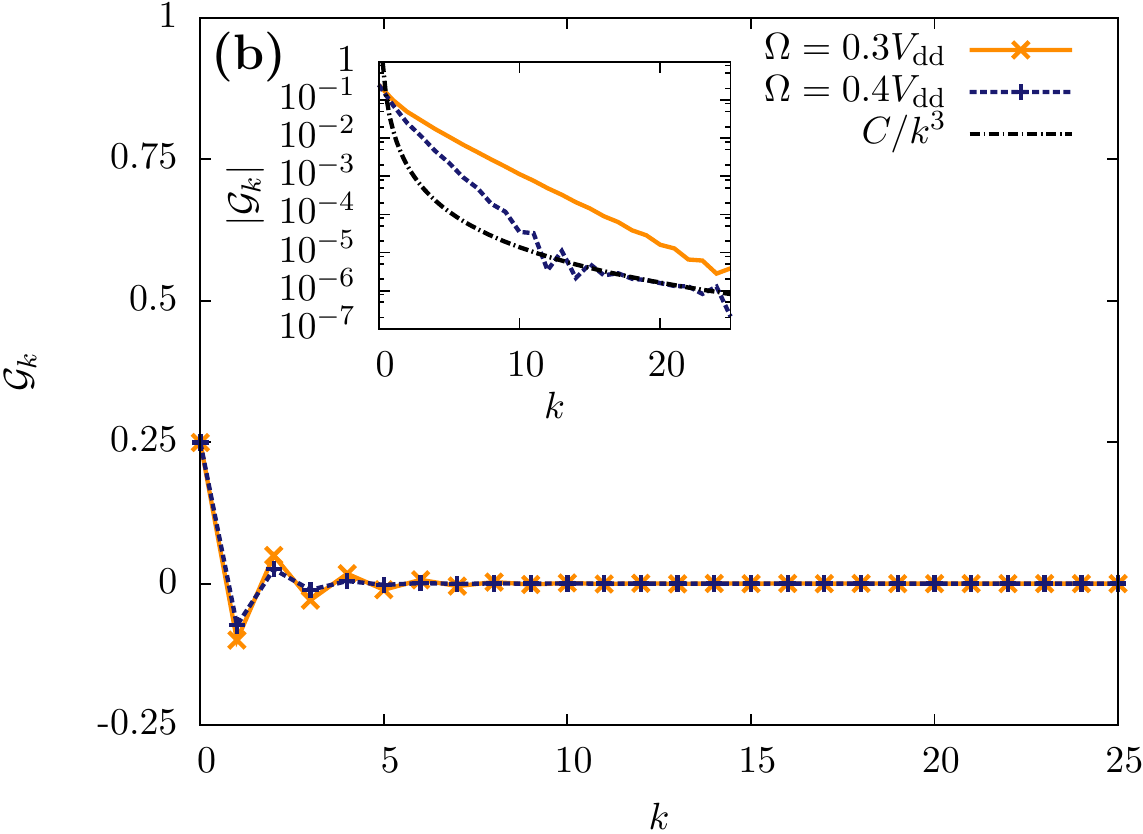}
  \caption{DDC functions $\mathcal{G}_k$ for the ground states of
    Hamiltonian (\ref{eq:hamiltonian}) for $\delta=1.1
    V_{\mathrm{dd}}$ and (a) $\Omega=0.1V_{\mathrm{dd}}$,
    $0.2V_{\mathrm{dd}}$, and (b) $\Omega=0.3V_{\mathrm{dd}}$,
    $0.4V_{\mathrm{dd}}$. The insets show the modulus
    $|\mathcal{G}_k|$ on a logarithmic scale. Here we show results for
    a system with $50$ sites and periodic boundary conditions. We find
    states with a marked crystalline structure of periodicity 2 in
    panel (a) and a phase transition at $0.2V_{\mathrm{dd}}< \Omega <
    0.3V_{\mathrm{dd}}$ in terms of the decay behaviour of
    $|\mathcal{G}_k|$. In contrast to panel (a), in (b)
    $|\mathcal{G}_k|$ decays exponentially. In the case of
    $\Omega=0.4V_{\mathrm{dd}}$ the long-range decay behaviour
    ($k\gtrsim 10$) follows a power-law decay $\propto Ck^{-3}$ (with
    $C \approx 1.7 \times 10^{-2}$).}\label{fig:ddc2ndtransition}
\end{figure}

In figure~\ref{fig:ddc2ndtransition}(a) we show a characteristic DDC in
the parameter regime where the filling of the lattice is approximately
$N/2$, choosing a detuning of $\delta=1.1 V_{\mathrm{dd}}$, with both
$\Omega=0.1 V_{\mathrm{dd}}$ and $\Omega=0.2V_{\mathrm{dd}}$. The
inset in figure~\ref{fig:ddc2ndtransition}(a) shows the modulus of the
correlation function, $| \mathcal{G}_k |$ on a logarithmic scale. In
the regime of the half filling plateau we indeed find crystalline
structure with a pronounced zigzag pattern. The long-range crystalline
order is indicated by the decay of $|\mathcal{G}_k |$ to a non-zero
value as a function of the site index $k$. We find that for increasing $N$
the value of the constant to which $|\mathcal{G}_k |$ decays 
becomes independent of the system size.

We note, already from the comparison in
figure~\ref{fig:ddc2ndtransition}(a) that as $\Omega/V_\mathrm{dd}$
increases the strength of the long range crystalline order
decreases. If we continue to increase $\Omega/V_\mathrm{dd}$, we find
that this order is broken entirely, as shown in
figure~\ref{fig:ddc2ndtransition}(b), with the same parameters, but
for Rabi frequencies $\Omega=0.3 V_\mathrm{dd}$ and $\Omega=0.4
V_\mathrm{dd}$. In these cases we also observe peaks with periodicity
2, however, these occur only for very small values of $k \lesssim
5$. As $k$ increases, $| \mathcal{G}_k |$ decays to zero, indicating
the breakdown of the long-range order. This appears as a phase
transition from crystalline order to essentially a ``paramagnetic
phase'', in which the correlations typically decay exponentially. Here
we note a special feature of the correlation functions due to the
long-range interactions, however, which is most clearly visible in the
results for $\Omega=0.4 V_{\mathrm{dd}}$ in
figure~\ref{fig:ddc2ndtransition}(b). There we observe that the
long-range order has been replaced by exponential decay at short
distances ($k \lesssim 10$), but that this changes from exponential to
a power law-decay. By fitting a function we find that in this region
$|\mathcal{G}_k| \approx C/k^3$, with a constant $C$. This is the same
power law decay as the interaction term, as is consistent with similar
results derived for other long-range spin models
\cite{Deng05,Schuch06}. Physically, the correlations initially decay
exponentially due to the primary influence on a given particle being
from its nearest neighbours (analogous to exponentially decaying
correlations in the paramagnetic phase of an Ising model). However,
once correlations arising from nearest neighbour interactions become
sufficiently small, long-distance correlations will be dominated by
effects of the direct interaction between distant particles. Thus,
long range correlations are expected to decay at large distances with
the same power as the interaction \cite{Schuch06}.

\begin{figure}[tb]
  \centering
  \includegraphics[width=0.45\textwidth]{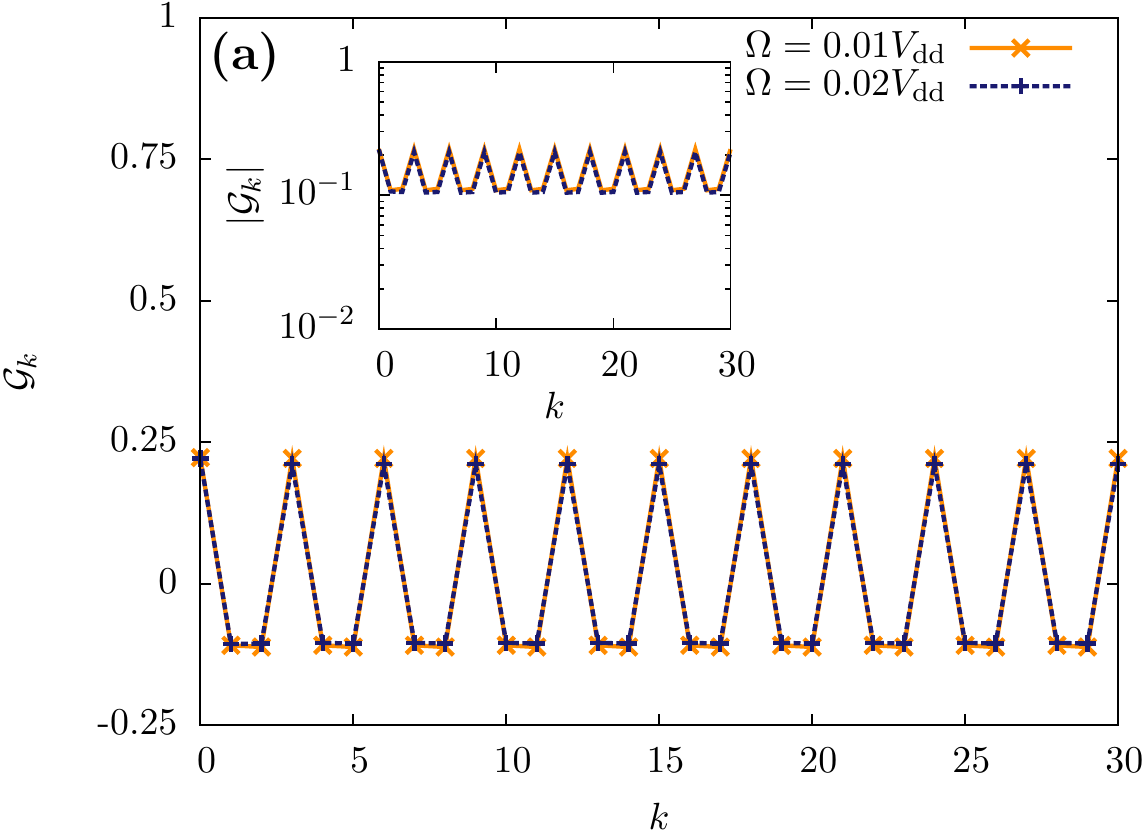}
  \includegraphics[width=0.45\textwidth]{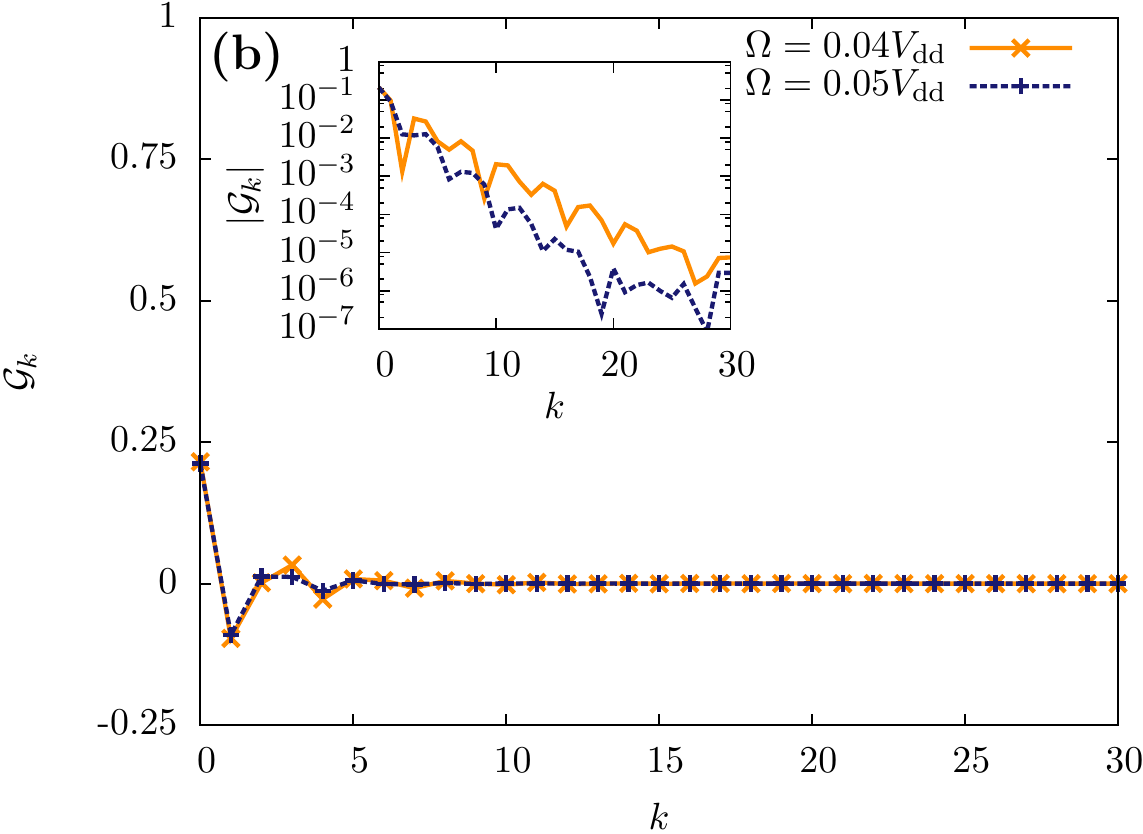}
  \caption{DDC functions $\mathcal{G}_k$ for the ground states of
    Hamiltonian (\ref{eq:hamiltonian}) for $\delta=0.25
    V_{\mathrm{dd}}$ and (a) $\Omega=0.01V_{\mathrm{dd}}$,
    $0.02V_{\mathrm{dd}}$, and (b) $\Omega=0.04V_{\mathrm{dd}}$,
    $0.05V_{\mathrm{dd}}$. The insets show the modulus
    $|\mathcal{G}_k|$ on a logarithmic scale. Here we show results for
    a system with $60$ sites and periodic boundary conditions. We find
    a marked crystalline structure in panel (a) with periodicity 3
    and a phase transition at around $\Omega \approx
    0.03V_{\mathrm{dd}}$ indicated by the exponential decay behaviour
    of $|\mathcal{G}_k|$ in panel (b).}\label{fig:ddc3rdtransition}
\end{figure}

We find similar behaviour for crystalline states at other densities,
including occupation periodicities larger than two. When the detuning
is decreased from $\delta \approx 0.5 V_{\mathrm{dd}}$ it becomes less
favourable to occupy $\ket{\beta}_k$ states in the system and therefore
in the case of small $\Omega$, crystalline phases with larger spacing
between these occupations appear in order to reduce the dipole-dipole
interaction energy. For example, in figure~\ref{fig:ddc3rdtransition} we
show the DDC indicating an occupation of the states $\ket{\beta}_k$ on
every third site for $\Omega \sim 10^{-2} V_{\mathrm{dd}}$ and $\delta
= 0.25 V_{\mathrm{dd}}$. To make this periodicity clearer, we present
results for a system with $N=60$ sites.  As in
figure~\ref{fig:ddc2ndtransition}, in figure~\ref{fig:ddc3rdtransition} we
observe the characteristic long range order in the DDC, combined with
a marked peak on every third lattice site. Again we find a phase
transition away from crystalline order, with an exponential decay of
$|\mathcal{G}_k|$ exhibited for larger values of $\Omega$. Note that
the crystalline phase with 1/3 filling occurs in a much smaller region
of parameter space than the crystalline phase with 1/2 filling.

In what follows, we are interested in time-dependent creation of these
crystalline states in the case of polar molecules or Rydberg atoms
loaded into an optical lattice. These systems will typically occupy $N
\sim 50$ lattice sites, with open boundary conditions generated by the
unoccupied ends of the chain. Thus, rather than attempting to derive
the full phase diagram for an infinite system, we focus on the case of
open boundary conditions relevant to the experiments. Although the
correlation functions are modified by the boundary conditions, we show
below that the characteristic signatures of different phases can still
be clearly identified.

\begin{figure}[htb]
  \centering
  \includegraphics[width=0.45\textwidth]{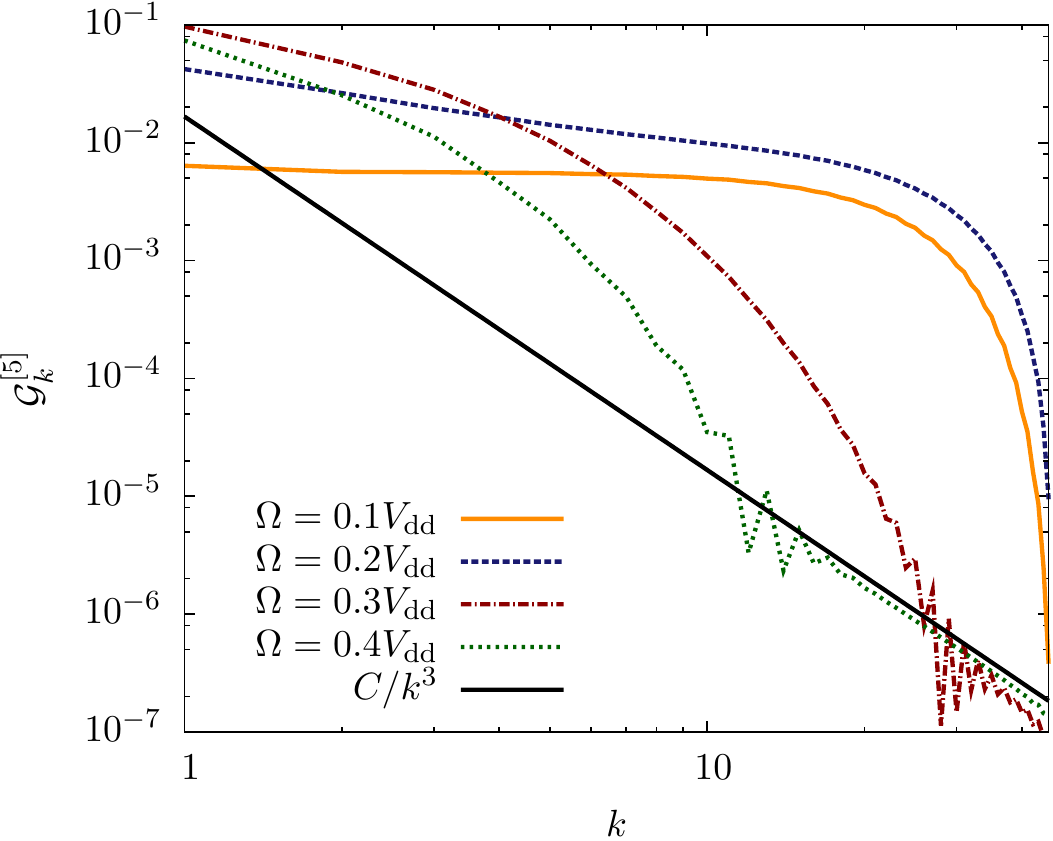}
  \caption{DDC functions evaluated from site index $i=5$ in a
    $50$-site system with open boundary conditions,
    $\mathcal{G}_k^{[5]}$. The results shown are for the same
    parameter regime as in figure~\ref{fig:ddc2ndtransition}. Despite
    strong boundary effects, a crystalline phase is still
    distinguishable from a phase with such structure via the behaviour
    over short ranges $k \lesssim 10$. In the crystalline phase the
    long-range decay follows a power law behaviour proportional to
    $Ck^{-3}$ ($C\approx 1.7 \times 10^{-2}$ is a
    constant).}\label{fig:ddc2ndbox}
\end{figure}

In figure~\ref{fig:ddc2ndbox} we show the DDC evaluated from site
number $5$ of a $N=50$ site system, i.e. $\mathcal{G}_k^{[5]}$ with
open boundary conditions. We choose the same parameters as were shown
in figures~\ref{fig:ddc2ndtransition}(a) and (b), including
crystalline phases with periodicity 2. Despite obvious effects of the
open boundary conditions we can clearly distinguish crystalline
behaviour ($\Omega=0.1V_{\mathrm{dd}}, 0.2V_{\mathrm{dd}}$) from the
case were exponentially decaying density-density correlations are
present ($\Omega=0.3V_{\mathrm{dd}}, 0.4V_{\mathrm{dd}}$). While in
the crystalline phase we find very little decay for increasing $k$
until the effects of the boundary become apparent ($k \gtrsim 20$), in
the case of $\Omega \gtrsim 0.3$ the correlations rapidly decay below
values of $10^{-5}$. As in figure~\ref{fig:ddc2ndtransition} we still
observe a long-range power law decay for $k \gtrsim 10$ proportional
to $k^{-3}$.

In relation to figure~\ref{fig:phase_diag_sketch}, the results
presented in figures~\ref{fig:ddc2ndtransition}-\ref{fig:ddc2ndbox}
give evidence for the crystalline phases at filling factors $1/2$ and
$1/3$ on the lattice, and show for each that there exist transitions as
$\Omega$ is increased to non-crystalline phases. This justifies that
the lobes are finite along the $\Omega$ axis, and shows that the
boundary depicted corresponds to a phase transition.

Note that the DDC functions can be measured
experimentally both for Rydberg atoms and for polar
molecules. Detection of these correlation functions is discussed in
more detail in section~\ref{sec:phys-impl-long}.

\section{Dynamical creation of crystalline states}
\label{sec:dynam-cryst-state}

\subsection{Time-dependent state preparation}

We now discuss the dynamical creation of crystalline states in our
system. The natural initial state for the experiments corresponds to
having all of the particles in the non-interacting states
$\ket{\alpha}_k$, which would be the ground states for neutral atoms,
or a single dressed rotational state for polar molecules. In terms of
the effective model, this state is the ground state of the Hamiltonian
(\ref{eq:hamiltonian}) for large negative detuning and small
$\Omega$. The question is now whether one can time-dependently vary
the laser parameters $\Omega$ and $\delta$, i.e. find a trajectory in
the $\Omega$-$\delta$-plane, so that this initial state is
adiabatically transferred to a crystalline state of the form discussed
in section~\ref{sec:ground-states-system}.

\begin{figure}[tb]
\centering
\includegraphics[width=0.45\textwidth]{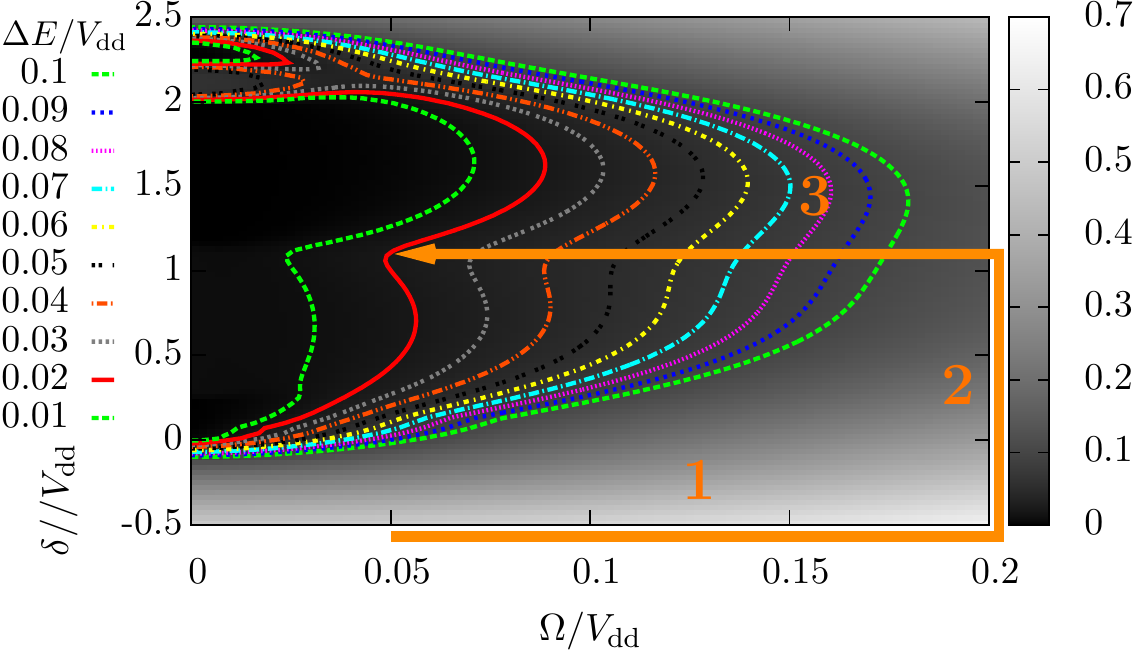}
\caption{Shaded plot of the energy gap $\Delta E$ between the ground
  and the first excited state as a function of $\Omega$ and
  $\delta$. The results are from exact calculations for an $8$-site
  system with Hamiltonian (\ref{eq:hamiltonian}). The lines indicate
  the contours of constant $\Delta E$. A path, which consists of three
  segments and yields an adiabatic passage to a crystalline ground
  state, is sketched as orange arrow (see text). }\label{fig:gap}
\end{figure}

In order to guide our discussion of this question, we first consider a
small model system consisting of $8$ sites, where exact
diagonalization of the Hamiltonian (\ref{eq:hamiltonian}) is
straightforward. We can then estimate possible paths for the adiabatic
transfer based on the size of the energy gap $\Delta E$ between the
ground state and excited states. During the adiabatic ramp, we would
like $\Delta E$ to remain as large as possible to suppress
non-adiabatic transitions to excited states of the effective model. In
figure~\ref{fig:gap} we show a shaded plot of the energy gap between
the ground state and the first excited state as a function of $\Omega$
and $\delta$, with lines of constant energy gap marked in the
plot. Close to $\Omega=0$ we find a large region where the energy gap
assumes small values and where one can recognize lobe-like
structures. Especially visible are lobes at values of $\delta \approx
2.25 V_\mathrm{dd}$. As mentioned before these correspond to states with
regular patterns of two subsquent sites being in the strongly
interacting state. In contrast to the lobe for filling fraction $N/3$
of the strongly interacting state at $\delta \approx 0.25 V_\mathrm{dd}$,
these are enhanced due to the larger increase in interaction energy
when adding single occupations of the strongly interacting state to
the system. Therefore, these lobes are more robust when increasing
$\Omega$. In general, the system is too small to observe lobes of
filling fractions $N/3$ and smaller and these are only visible as
small kinks for $\delta \approx 0.25 V_\mathrm{dd}$. With increasing Rabi
frequency $\Omega$ the energy gap increases. The requirement of having
at any instance of time a sufficiently large energy gap is most easily
fulfilled if one chooses to ``steer'' a path in parameter space around
those regions where the gap is the smallest. A simple path that seems
to achieve this is, for example, to first increase $\Omega$ at
constant large negative detuning $\delta$ and then increase $\delta$
at a constant large $\Omega$. Both of these two steps can be achieved
by changing the parameters $\delta$ and $\Omega$ rapidly, due to the
large gap $\Delta E$ present at all times.  Afterwards, $\Omega$ can
be decreased to a small value on a line of constant $\delta$.  This
path consisting of these three segments is marked on
figure~\ref{fig:gap} by an orange arrow.

\begin{figure}[htb]
\centering
\includegraphics[width=0.45\textwidth]{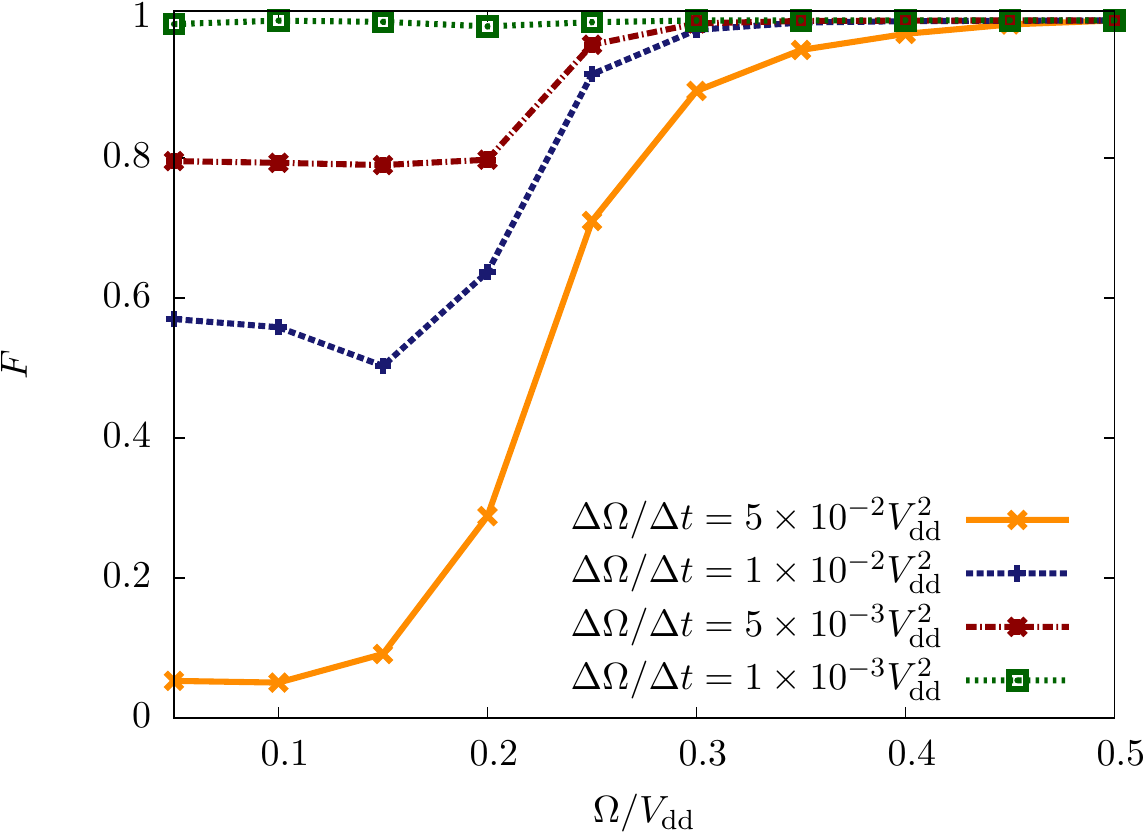}
\caption{The fidelity of the adiabatically evolved state with the
  current ground state, when changing $\Omega$ from $0.50
  V_{\mathrm{dd}}$ to $0.05 V_{\mathrm{dd}}$ at constant $\delta=1.1
  V_{\mathrm{dd}}$ in segment 3 of the adiabatic preparation
  path. Results are shown for different variation rates $\Delta \Omega
  /\Delta t$ within a 30-site system with Hamiltonian
  (\ref{eq:hamiltonian}). For decreasing variation rates the final
  fidelity for the crystalline state (at $\Omega=0.05 V_\mathrm{dd}$)
  increases towards 1.}\label{fig:fidelity}
\end{figure}

Now we would like to test these insights by investigating the
time-dependent propagation of the system along this path in parameter
space. In order to investigate the scaling with the system size, we
study systems of size $20$, $30$, $40$ and $50$ lattice sites. We note
that the large sizes here are already typical of likely experimental
setups. As an example, we choose to prepare a crystalline state with
excitation periodicity 2, which is the ground state of the our model
for the parameters $\Omega=0.05 V_{\mathrm{dd}}$ and $\delta=1.1
V_{\mathrm{dd}}$. We perform real time simulations for a path
consisting of three segments, as sketched in figure~\ref{fig:gap}.  We
start at $\Omega=0.05 V_{\mathrm{dd}}$ and $\delta=-3
V_{\mathrm{dd}}$. There, the inner product between the ground state of
the Hamiltonian and the state in which all atoms are in the
$\ket{\alpha}_k$ state is sufficiently large for all system sizes. The
segments of the path are:
\begin{enumerate}
\item We increase $\Omega$ from $0.05 V_{\mathrm{dd}}$ to $0.5
  V_{\mathrm{dd}}$ at $\delta=-3$ with a variation rate $\Delta \Omega
  / \Delta t = 0.05 V_{\mathrm{dd}}^2$.
\item We increase $\delta$ from $-3 V_{\mathrm{dd}}$ to $1.1
  V_{\mathrm{dd}}$ at $\Omega=0.5 V_{\mathrm{dd}}$ with a variation
  rate $\Delta \delta / \Delta t = 0.05 V_{\mathrm{dd}}^2$.
\item We decrease $\Omega$ from $0.5 V_{\mathrm{dd}}$ to $0.05
  V_{\mathrm{dd}}$ with several variation rates $\Delta\Omega /\Delta
  t$.
\end{enumerate}
Due to the large gaps in the energy spectrum it is possible for
segments 1 and 2 to obtain states with a high fidelity $F \equiv |
\braket{\Psi_G (\Omega, \delta)}{\Psi(t)} |^2 \sim 1$ in the
comparison of our evolved state, $\ket{\Psi(t)}$, with the ground
state of the effective model with the corresponding parameters,
$\ket{\Psi_G (\Omega, \delta)}$ on that path.  The most sensitive part
of this process is the third segment. In figure~\ref{fig:fidelity} we
show the fidelity $F$ of the adiabatically evolved state along segment
$3$ as a function of $\Omega(t)$ and for various variation rates
$\Delta \Omega / \Delta t$. On $N=30$ lattice sites, we find that for
the three rates $\Delta \Omega / \Delta t = 5 \times 10^{-2}
V_\mathrm{dd}, 1 \times 10^{-2} V_\mathrm{dd},$ and $5 \times 10^{-3}
V_\mathrm{dd}$, this fidelity drops below values of $80\%$, which
occurs when crossing the phase transition from
figure~\ref{fig:ddc2ndtransition}, i.e. in the regime
$0.2V_{\mathrm{dd}}< \Omega < 0.3V_{\mathrm{dd}}$. At this point the
ground state gap becomes too small and excitations to higher excited
states analogous to Landau-Zener tunnelling processes occur. However
when using a rate of $\Delta \Omega / \Delta t = 1 \times 10^{-3}
V_\mathrm{dd}$ we find that the fraction of the state that is lost
into higher excitations remains reasonably small and the final
fidelity with the crystalline state at $\Omega=0.05 V_{\mathrm{dd}}$
is larger than $F\approx 99\%$. Thus, we have shown that a crystalline
state can be produced with very high fidelity in a finite system of
$30$ sites. The timescale which is required for all three segments is
of the order of $500/V_\mathrm{dd}$, and we will show that this is
experimentally realisable for our two physical implementations below.

In an experiment with either polar molecules or neutral atoms coupled
to excited Rydberg states in an optical lattice, the final state could
be characterized by measuring DDC functions. These can be most easily
detected directly by {\em in situ} imaging experiments \cite{Bakr10}. In the
case of polar molecules, the crystalline structure will be visible in
noise-correlation measurements of state-selective momentum
distributions of molecules released from the lattice \cite{Altmann04,
  Foelling05, Greiner05}. In these correlation measurements, the
periodicity of the crystalline structure would be present as a peak at
the corresponding reciprocal lattice vector. In the case of Rydberg
excitations, direct measurement of DDC
functions could also be made by imaging the sample on a channel plate
detector \cite{Pohl10}.

\begin{figure}[htb]
\centering
\includegraphics[width=0.45\textwidth]{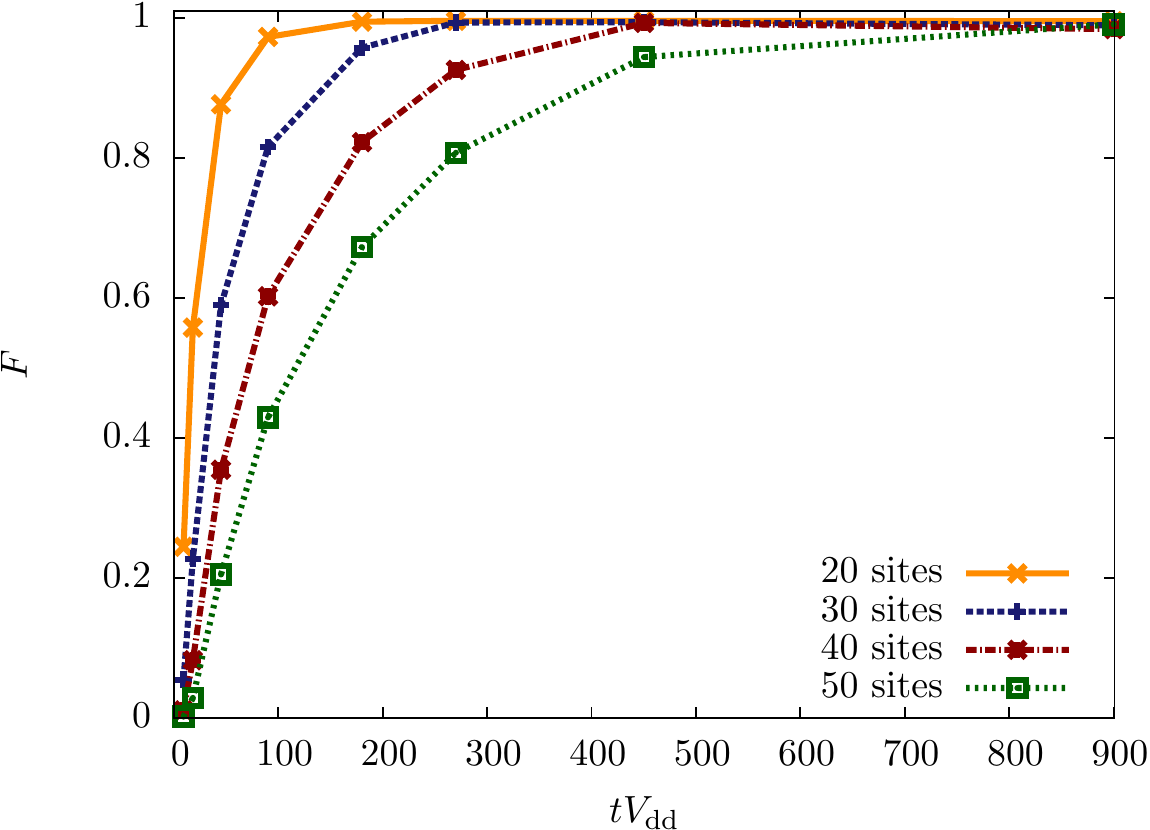}
\caption{The fidelity of the time-evolved state with the target
  crystalline state, at $\Omega=0.05 V_\mathrm{dd}$, and $\delta=1.1
  V_\mathrm{dd}$ as a function of the time required for the segment
  3. Here we show results for several system sizes of 20, 30, 40 and
  50 lattice sites. This fidelity decreases with the increasing system
  size and with decreasing preparation time.}\label{fig:fidelity_M}
\end{figure}

An important question is: to what extent the same fidelities we have
observed for a system size of 30 lattice sites can also be achieved
for larger system sizes. In general we expect the energy gap between
ground and excited states to become smaller with increasing system
size, and we expect the measure of the fidelity to become more
sensitive due to the exponentially increasing size of the Hilbert
space. To analyse how much this affects the final fidelity, in
figure~\ref{fig:fidelity_M} we plot the fidelity achieved after ramping
the laser parameters through all segments of our preparation
procedure, as a function of the time required for the final segment
$3$. We consider system sizes of $20, 30, 40$ and $50$ sites.

We find that for all system sizes the fidelity is $> 94 \%$ when
choosing a variation rate slower than $\Delta \Omega/ \Delta t=1\times
10^{-3} V_{\mathrm{dd}}^2$, which corresponds to an operation time of
approximately $450 / V_{\mathrm{dd}}$ on segment 3. In the case of an
even slower rate of $\Delta \Omega/ \Delta t=5\times 10^{-4}
V_{\mathrm{dd}}^2$ which requires a time of approximately
$900/V_\mathrm{dd}$, the fidelity becomes larger than $98 \%$, even in
the case of a system with $50$ lattice sites. This fidelity is very
high, especially given that the Hilbert space dimension is $2^{50}$ in
the latter case.

\subsection{Comparison with experimental parameters}

Let us compare the time required to reach a final crystalline phase
with experimentally accessible timescales, which are in general
limited by the finite lifetime of the strongly interacting state. In
the scheme presented above, at the slowest variation rate of $\Delta
\Omega/ \Delta t=5\times 10^{-4} V_{\mathrm{dd}}^2$ on segment 3, a
total operation time of the order $1000 / V_\mathrm{dd}$ is required.
Note that this time can be further dramatically decreased by choosing
shorter paths or by not utilising a constant ramping rate but rather
reducing it with time while approaching the critical point of the
phase transition.

In the case of the polar molecule implementation (see
section~\ref{sec:polar-molecules-an} for more details), typical
parameters give $V_\mathrm{dd} \approx 10 \hbar \, \mathrm{kHz}$,
assuming a permanent dipole moment $d_0=1 \, \mathrm{Debye}$ and a
lattice spacing of $a=400\,\mathrm{nm}$. This translates to a time for
the complete preparation of approximately $0.1\, \mathrm{s}$, which is
significantly shorter than recently measured lifetimes of molecules in
an optical lattice of $8\,\mathrm{s}$ \cite{Danzl09}. By choosing
polar molecules with larger dipole moments like LiCs ($d_0 \approx 5.5
\, \mathrm{Debye}$ \cite{Carr09}), the preparation time can be further
reduced down to the order of a few $\mathrm{ms}$.

In the case of the Rydberg atom implementation (see
section~\ref{sec:rydb-excit-an} for more details), if we choose a
principal quantum number $n=14$, then $V_{\mathrm{dd}}\approx 7 \hbar
\, \mathrm{GHz}$ (assuming a lattice spacing of
$a=400\,\mbox{nm}$). Hence, the total operation time corresponds to
approximately $140 \ \mathrm{n s}$ in the experiment.  This value is
well below the typical lifetime of a $n=14$ Rydberg level of Lithium,
which can be found to be approximately $2.2\, \mathrm{\mu s}$
\cite{Theodosiou84}. Note that these estimates neglect the effects of
blackbody radiation, which can redistribute population amongst
excited levels. For Rydberg atoms, in general the lifetime increases
with increasing $n$ and furthermore the required operation time
decreases as $n^{-4}$.  However, the disadvantage of large principal
quantum numbers is the small level distances $\Delta_{\mathrm{inter}}$
and $\Delta_{\mathrm{intra}}$ (see figure~\ref{fig:level_structure}),
which both decrease with $n$. This limits the experimentally
accessible values of $\delta$. In the case $n=14$ ($a=400\,\mbox{nm}$)
at an electric field strength of $F_\mathrm{el}=100\,\mbox{kV/m}$, we
can estimate by using the results known from the hydrogen atom
$\Delta_\mathrm{inter} \approx 4.6 \times 10^3 V_{\mathrm{dd}}$ and
$\Delta_\mathrm{intra} \approx 90 V_{\mathrm{dd}}$. The field strength
is thereby well below the Inglis-Teller limit, which is in this case
$319\,\mbox{kV/m}$ \cite{Gallagher94}. This justifies the validity of
our two-level model, derived in section~\ref{sec:rydb-excit-an}, in
this regime.

This comparison to the lifetime of a single Rydberg atom is
oversimplified, as in the presence of $N_\beta$ excited atoms, the
rate of single decay events will increase roughly proportional to
$N_\beta$ \cite{footnote2}. However, as the
system size (and hence $N_\beta$) becomes larger, the effect of a
single decay on the crystalline structure should be reduced. Whilst
the many-body state fidelity will be very sensitive to a single decay,
DDC functions should be robust, at least on
length scales smaller than those separating places where spontaneous
emissions have occurred. The effect on the many-body state of single
decay events also depends on when during the excitation process they
occur. A full treatment of this excitation process including decay
events could be achieved by studying the dynamics of a master equation
including spontaneous emission events.

Therefore and from the results in figure~\ref{fig:fidelity_M} we
conclude that the adiabatic passage should be experimentally
realisable for both our implementations of the effective spin-model
with a size of $\sim 50$ sites. For larger system sizes, longer
timescales will be required to reach as high fidelities as we have
obtained here. However, the properties of the final states will
typically be determined in an experiment by measurements of the
correlation functions. In contrast to the state fidelity, correlation
functions (together with the physical character of the state) become
more robust to small non-adiabaticities (especially those resulting in
localized defects) in larger systems.

\section{Physical implementations of a long-range spin model}
\label{sec:phys-impl-long}

\subsection{Polar molecules in an optical lattice}
\label{sec:polar-molecules-an}

In the following we describe how to implement Hamiltonian
(\ref{eq:hamiltonian}) with a system of polar molecules. We consider
$N$ molecules confined to the lowest motional level on subsequent
sites of a deep 1D optical lattice, with modefunction
$\phi_k(\mathbf{x})$ on site $k$. We assume that the lattice is
sufficiently deep that tunnelling between wells of the lattice is
strongly suppressed on typical experimental timescales.  In particular
we consider molecules in their electronic vibrational ground-state
manifold with a closed shell structure $^1\Sigma$, which possess
permanent dipole moments (e.g. $^{40}$K$^{87}$Rb like in recent
experiments \cite{Ni08}). The setup is depicted in
figure~\ref{fig:pmsetup_1}. We use a strong and a weak microwave field
linearly polarized along the $z$-axis and an additional static field
aligned along the same direction. Given the fact that we consider
samples of the order of a few to hundred $\mathrm{\mu m}$ one can
neglect the spatial variation of the respective static and microwave
field along the optical lattice axis, which is aligned in the
$\mathbf{e}_x$-direction.

The low energy spectrum of a single molecule at site $k$ in our setup
is well described by rotational excitations, governed by a rigid
rotor Hamiltonian coupled to the fields via an electric dipole
interaction \cite{Buechler07, Micheli07, Herzberg50, Brown03}
\begin{align} \label{eqn:ham_single_molecule} 
  \hat H_{\mathrm{rot}}^k(t)
  =
  B \hat {\bf J}_k^2
  -{\hat d}^z_k \left(
  {E_\mathrm{dc}} +
  {E_1^z}(t) +
  {E_2^z}(t) \right).
\end{align}
Here, $B$ is the rotational constant typically of the order of tens of
GHz ($\hbar\equiv 1$) \cite{diatomicdatabase}, $\hat {\bf J}_k $ is
the angular momentum operator, $\hat d^z_k$ the $z$-component of the
dipole operator and ${E_\mathrm{dc}}$,
${E_1^z(t)}=E^z_1\exp(-i\omega_1t)+\mathrm{c.c.}$, and
${E_2^z(t)}=E^z_2\exp(-i\omega_2t)+\mathrm{c.c.}$ the static, weak and
strong external microwave fields, respectively.

The eigenstates of $\hat {\bf J}_k^2$ are denoted $\ket{J,m_J}_k$ with
eigenvalues $J(J+1)$ $(\hbar\equiv 1)$, and $m_J=-J, \dots, J$ being
the eigenvalues of the projection onto the $z$-axis, $\hat
{J}_k^z$. The dipole operator $\hat d_k^z$ couples rotational states
with differences $\Delta J=\pm 1$, and we recall that the eigenstates
of equation~(\ref{eqn:ham_single_molecule}) have no net dipole moment. The
non-vanishing elements of the operator $\hat d_k^z$ are given by
$\bra{J\pm 1,m_J}\hat d_k^z \ket{J,m_J}_k=d_0( J , m_J ; 1, 0 | J \pm
1, m_J) ( J , 0 ;1, 0 | J \pm 1, 0) \sqrt{(2J+1)/(2( J \pm 1) + 1)}$
in terms of Clebsch-Gor\-dan coefficients $(J_1, m_1; J_2, m_2 | J, m)$,
where $d_0$ denotes the permanent electric dipole moment along the
internuclear axis, which is typically of the order of a few Debyes
\cite{Krems09, Carr09}.

By applying electric fields one lifts the degeneracy of the state
manifolds with $J>0$ and induces a finite dipole moment in the dressed
rotational states. Since the fields are aligned along the quantisation
axis, $m_J$ is conserved, and we focus on states with $m_J=0$ in our
setup, which are connected to the ground state. The induced dipole
moments of two molecules at distinct sites $k$ and $l$ then lead to an
energy shift due to the dipole-dipole interaction potential. 

In order to describe this, we need to consider many molecules on the
lattice. Molecules in an internal state denoted $\sigma$ can then be
represented by field operators $\hat \Psi_\sigma (\mathbf{x})$,
obeying the standard bosonic or fermionic commutation relations. We
consider a situation where no motional modes other than
$\phi_k(\mathbf{x})$ for each site $k$ are populated during the
experiment, which is valid provided that the band separation
$\omega_T$ is substantially larger than the laser coupling parameters
and interaction energies.  Thus, we can expand the field operators as
\begin{align}
  \hat \Psi_\sigma(\mathbf{x})&=\sum_k \phi_k(\mathbf{x}) \hat g_k^{\sigma}
\end{align}
and identify individual molecules occupying each mode described by the
operators $\hat g_k^{\sigma}$.  Since all rotational states we
consider here have $m_J=0$, they are only coupled by an interaction
proportional to $\hat d^z_k \hat d^z_l$, which can be integrated over
the modefunctions in each lattice site. The resulting many-body
Hamiltonian of a system for $N$ molecules can be written as
\begin{align} \label{eqn:ham_lattice_molecules} 
  \hat H_{\mathrm{pm}}(t)
  \approx
  \sum_{k}^N \hat H_{\mathrm{rot}}^k (t)
  + V_{\rm dd}^{(1)} \sum_{k\neq l}^N \frac{\hat d^z_k \hat d^z_l }{|k-l|^3}
  ,
\end{align}
with lattice spacing $a$, and where $V_{\rm dd}^{(1)} \approx 1/(8 \pi
\epsilon_0 a^3)$. Note that we neglected corrections to the shape of
the interactions, which are of the order
$\order{l_\mathrm{H}/(a|k-l|)}$, where $l_\mathrm{H}$ denotes the
characteristic harmonic oscillator length for the optical potential
well on each site.

A suitable choice of external electric fields then makes it possible
to strongly couple/dress locally two (effective) rotational levels,
such that only one level has a strong induced dipole moment, while at
the same time the transition dipole moment remains negligibly
small. This results in dipole-dipole interactions in
equation~(\ref{eqn:ham_lattice_molecules}) reducing to the diagonal form of
equation.~(\ref{eq:hamiltonian}).

\begin{figure}[tb]
  \centering
 \includegraphics[width=0.45\textwidth]{Fig11}
 \caption{\label{fig:pmsetup}Level scheme for a system of polar
   molecules in a combination of static and microwave fields,
   cf. figure~\ref{fig:pmsetup_1}. (a) Energy levels $E_J$ of a molecule
   in a static electric field $E_{\rm dc}$. The field orients the
   dipoles, i.e. induces finite permanent dipole moments in each
   rotational state $|J\rangle_{\rm dc}$. (b) A strong
   linearly-polarized microwave field $E^z_2(t)$ couples the states
   $|1\rangle_{\rm dc}$ and $|2\rangle_{\rm dc}$ the Rabi-frequency
   $\Omega_2$ and detuning $\Delta_2$. Fine-tuning the microwave-field
   (with respect to the static field) makes it possible to cancel the
   permanent dipole-moment in one of the two dressed states,
   $|\alpha_+\rangle$ and $|\alpha_-\rangle$,
   cf. $d_\alpha\rightarrow0$. A second (weaker) microwave field
   $E^z_1(t)$ finally couples the ground state
   $|\beta\rangle\equiv|0\rangle_{\rm dc}$ to the dressed state with
   vanishing dipole $|\alpha\rangle$ with an effective Rabi-frequency
   $\Omega_1$ and detuning $\Delta_1$.}
\end{figure}

We now detail our appropriate choice of the electric fields to obtain
the desired dressed states. The key idea is to apply the static field
$E_\mathrm{dc}$ to induce a finite dipole moment in the rotational
states, leading to the dressed states $\ket{J}_{\mathrm{dc}}$, defined
by $[ B \hat {\bf J}_k^2 - \hat d_k^zE_\mathrm{dc} ]
\ket{J}_{\mathrm{dc}}=E_J \ket{J}_{\mathrm{dc}}$ (see
figure~\ref{fig:pmsetup}(b), and to then couple the two states
$\ket{1}_{\mathrm{dc}}$ and $\ket{2}_{\mathrm{dc}}$ by a strong
microwave field $E^z_2(t)$, such that the dipole moment vanishes in
one of the resulting dressed states. The field $E^z_2(t)$ is tuned to
the transition between the two states with Rabi frequency
$\Omega_2\equiv \bra{2} \hat d_k^z \ket{1}_\mathrm{dc} E_2^z$ and
detuning $\Delta_2\equiv \omega_2 - (E_{2} - E_{1}) $ (see
figure~\ref{fig:pmsetup}). Diagonalizing the corresponding two-level
Hamiltonian and making the rotating wave approximation leads to the
two dressed states $\ket{\alpha_\pm}_k= \cos(\theta/2)
\ket{1}_{\mathrm{dc}} - \sin(\theta/2) \ket{2}_{\mathrm{dc}}$ with the
dressing angle $\theta\equiv \arctan(-2\Omega_2/\Delta_2)$ ($0 <
\theta < 2\pi$), where the state $\ket{\alpha_-}_k$
($\ket{\alpha_+}_k$) corresponds to $\theta < \pi$ ($\theta >
\pi$). The corresponding energies of the dressed states are
$E_\alpha^\pm=E_1-\Delta_2/2 \pm \sqrt{\Delta_2^2 + 4\Omega_2^2}/2$.

Finally, we couple the state $\ket{\alpha}_k \equiv \ket{\alpha_-}_k$
and the state $\ket{\beta}_k \equiv \ket{0}_{\mathrm{dc}}$ by the weak
electric field with Rabi frequency $\Omega_1 \equiv \bra{\alpha} \hat
d_k^z \ket{\beta}_k E_1^z$ and detuning $\Delta_1\equiv \omega_1 -
(E_\alpha-E_{0})$.  Thereby we assume that the field $E^z_1(t)$ is
much weaker than $E^z_2(t)$, so that the states $\ket{\alpha}$ and
$\ket{\beta}$ are unaffected from $E^z_1(t)$.  Going to the rotating
frame and making the rotating wave approximation the many-body
Hamiltonian becomes time-independent and reads
\begin{align} \label{eqn:ham_pm_pre} 
  \hat H_{\mathrm{pm}}
  =
  \Omega_1 \sum_k^N \hat \sigma_k^x
  -
  \Delta_1 \sum_k^N \hat n_k
  + V_{\rm dd}^{(1)} \sum_{k\neq l}^N \frac{\hat V_\mathrm{d}}{|k-l|^3}
  ,
\end{align}
with $\hat \sigma_x \equiv \ket{\beta}\bra{\alpha}_k +
\ket{\alpha}\bra{\beta}_k$ and $\hat n_k \equiv
\ket{\beta}\bra{\beta}_k $. In the last term of
equation~(\ref{eqn:ham_pm_pre}), $\hat V_\mathrm{d}$ is the
dipole-dipole interaction matrix in the rotating frame, which in the
two-molecule basis $\{ \ket{\alpha}_k\ket{\alpha}_l,
\ket{\alpha}_k\ket{\beta}_l, \ket{\beta}_k\ket{\alpha}_l,
\ket{\beta}_k\ket{\beta}_l \}$ is
\begin{align} \label{eqn:dd_matrix}
\hat V_\mathrm{d} 
\equiv
\left(
\begin{array}{cccc}
 v_{\alpha \alpha}^{\alpha \alpha} & 0 & 0 & 0   \\
 0 & v_{\alpha \beta}^{\alpha \beta} &  v_{\alpha \beta}^{\beta \alpha} & 0 \\
 0 &  v_{\alpha \beta}^{\beta \alpha} &  v_{\alpha \beta}^{\alpha \beta} & 0 \\
 0 & 0 & 0 &  v_{\beta \beta}^{\beta \beta} \\
\end{array}
\right)
.
\end{align}
Here, the non-vanishing elements of $\hat V_\mathrm{d}$ can be
evaluated in the rotating wave approximation from the bare
single-particle dipole-operator elements $d_{ij}\equiv \bra{i} \hat
d_k^z \ket{j}_\mathrm{dc}$:
\begin{align}
 v_{\alpha \alpha}^{\alpha \alpha } 
 &= 
 \left(
   \cos^2(\theta/2) d_{11} + \sin^2(\theta/2) d_{22}
   \right)^2 \nonumber \\
& \qquad+2 \sin^2(\theta/2) \cos^2(\theta/2) d_{12}^2 \label{eq:dd-elements-aaaa}\\
   v_{\alpha \beta}^{\alpha \beta}
  &= d_{00}
  \left(
    \cos^2(\theta/2) d_{11} + \sin^2(\theta/2) d_{22}
    \right) \\
  v_{\alpha \beta}^{\beta \alpha} 
  &= \cos^2(\theta/2) d_{10}^2 + \sin^2(\theta/2) d_{02}^2 \label{eq:dd-elements-abba}\\
  v_{\beta \beta}^{\beta \beta} 
  &= d_{00}^2
  .
\end{align}

Our goal is to find optimal values of the microwave dressing angle
$\theta$ and static field strength $E_\mathrm{dc}$ so that the desired
Hamiltonian [equation (1)] is realized with as small an error as
possible. Though from equation~(\ref{eq:dd-elements-abba}) we see that
there is no parameter choice where $v_{\alpha \beta}^{\beta \alpha}$
is exactly zero, we can find regimes where the model is implemented up
to very small errors. At this point we note that subtracting a
multiple of the identity from $V_{d}$ amounts to an overall shift of
the zero of energy, and thus we can equivalently write
\begin{align} \label{eqn:dd_matrix_subs}
\hat V'_\mathrm{d} 
\equiv
\left(
\begin{array}{cccc}
 v_{\alpha \alpha}^{\alpha \alpha} - v_{\alpha \beta}^{\alpha \beta}& 0 & 0 & 0   \\
 0 & 0 &  v_{\alpha \beta}^{\beta \alpha} & 0 \\
 0 &  v_{\alpha \beta}^{\beta \alpha} &  0 & 0 \\
 0 & 0 & 0 &  \delta v \\
\end{array}
\right)
,
\end{align}
where $\delta v \equiv v_{\beta \beta}^{\beta \beta} - v_{\alpha
  \beta}^{\alpha \beta}$.

\begin{figure}[htb]
  \centering
  \includegraphics[width=0.45\textwidth]{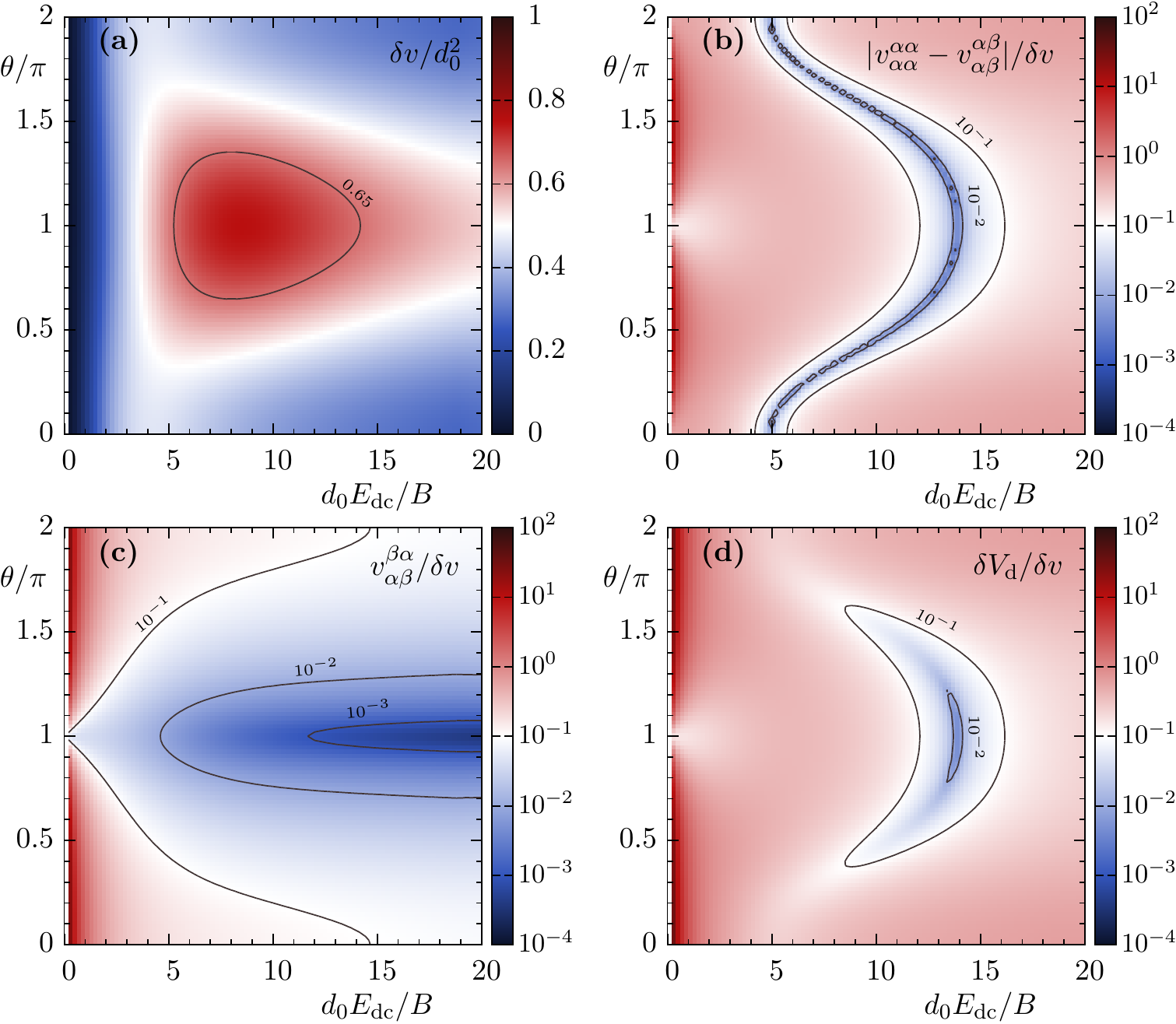}
  \caption{ Matrix elements of the dipole-dipole interaction matrix
    $\hat V_d$ in static and microwave fields, $E_\mathrm{dc}$ and
    $E_2^z(t)$. (a) Interaction strength for the strongly interacting
    state, $\delta v$ (in units of $d_0^2$). (b) The ratio $|
    v_{\alpha \alpha}^{\alpha \alpha} - v_{\alpha \beta}^{\alpha
      \beta} | / \delta v$ of the diagonal element of $\hat V_d$ and
    (c) of the off-diagonal element, $v_{\alpha \beta}^{\beta \alpha}/
    \delta v$, as a function of the static-field strength $d_0E_{\rm
      dc}/B$ and the microwave-dressing angle $\theta$ on a
    logarithmic scale. (d) The relative deviation, $\delta V_{\rm
      d}/ \delta v $, of the induced dipole-dipole
    coupling with respect to the model interaction as a function of
    $d_0E_{\rm dc}/B$ and $\theta$ on a logarithmic scale.  The solid
    lines show contours of constant value.
    \label{fig:pm_dipcontour}}
\end{figure}

In Figure \ref{fig:pm_dipcontour} we plot the matrix elements $\delta
v $, $| v_{\alpha \alpha}^{\alpha \alpha } - v_{\alpha \beta}^{\alpha
  \beta} | $, and $v_{\alpha \beta}^{\beta \alpha} $ as a function of
the microwave field angle $\theta$ and the electric field strengths up
to $E_\mathrm{dc}=20 d_0/B$. In figure~\ref{fig:pm_dipcontour}(a) we
show the magnitude of the interaction energy for the strongly
interacting state, $\delta v$. Figure~\ref{fig:pm_dipcontour}(b), and
figure \ref{fig:pm_dipcontour}(c) show the ratio of the other
interaction matrix elements to $\delta v$, i.e. $ | v_{\alpha
  \alpha}^{\alpha \alpha } - v_{\alpha \beta}^{\alpha \beta}| / \delta
v$, and $|v_{\alpha \beta}^{\beta \alpha}|/\delta v$. In panel (d) we
plot the (normalized) total deviation from an exact implementation of
the Hamiltonian [equation (1)]. This is quantified as $\delta
V_\mathrm{d} / \delta v$, given by
\begin{align}
  \delta V_\mathrm{d}  &\equiv
  \| \hat V'_\mathrm{d} -
  \delta v  
  \ket{\beta}_k\ket{\beta}_l\bra{\beta}_k\bra{\beta}_l \| \nonumber \\
  &=
  \sqrt{
    (v_{\alpha \alpha}^{\alpha\alpha} - v_{\alpha \beta}^{\alpha \beta} ) ^2
    +
    2  (v_{\alpha \beta}^{\beta \alpha})^2
  }.
\end{align}
We find that for field strengths $E_\mathrm{dc} \sim 14 B/d_0$, fine
tuning the microwave field at $\theta \sim 0.9 \pi$ makes it possible
to obtain deviations $\delta V_\mathrm{d} \sim 10^{-3} \delta v$,
while the interaction strength for the strongly interacting state is
$\delta v \approx 0.65 d_0^2$ (see contour line in panel (a) ). Thus
within small errors we obtain the model interaction with a typical
timescale of $V_{\mathrm{dd}} \equiv 2 \delta v
V_\mathrm{dd}^{(1)}\approx \delta v /4\pi \epsilon_0 a^3 \approx 10
\hbar \, \mathrm{kHz}$ for typical parameters ($a=400 \, \mathrm{nm}$,
$d_0=0.8 \, \mathrm{Debye}$).  An alternative approach using
circularly polarized coupling fields is also possible, which can give
rise to an implementation of the model for appropriately tuned
parameters \cite{gorshkov}.

\subsection{Neutral atoms coupled to Rydberg states}
\label{sec:rydb-excit-an}

In this section we show how to implement spin Hamiltonian
(\ref{eq:hamiltonian}) by making use of neutral atoms coupled to
excited Rydberg states. We consider alkali atoms confined in an
optical lattice in a single tube, with inter-site distance $a$
oriented along the $x$-direction.  Perpendicular to the lattice a
homogeneous electric field, $\mathbf{F}=F_\mathrm{el}\mathbf{e}_z$ is
applied, together with a laser which couples the ground state
$\left|g\right>$ of each atom to a specific chosen Rydberg-Stark
state.

\subsubsection{Dipole-Dipole Interaction}

While the atomic ground state is only slightly shifted by the external
electric field, the level structure of high-lying Rydberg states is
strongly altered compared to the field-free case. Nevertheless, the
azimuthal symmetry of the atomic Hamiltonian is preserved such that
$m$ (the quantum number of the $z$-component of the electronic orbital
angular momentum $L_z$) remains a good quantum number
\cite{Gallagher94}. States with large quantum defect, i.e. usually s-
and p-states, sustain a second-order Stark shift while the degenerate
states with higher angular momentum exhibit a linear Stark effect,
reminiscent of the hydrogen atom \cite{Zimmermann79}. The derivative
of the energy with respect to the electric field strength directly
yields the electric dipole moment of the given state. While the
precise level structure and therefore the dipole moment of the Stark
states depends on the element under consideration the exact results
known from the hydrogen atom usually provide good estimates for the
quantities of interest.

\begin{figure}[htb]
\centering
\includegraphics[width=0.45\textwidth]{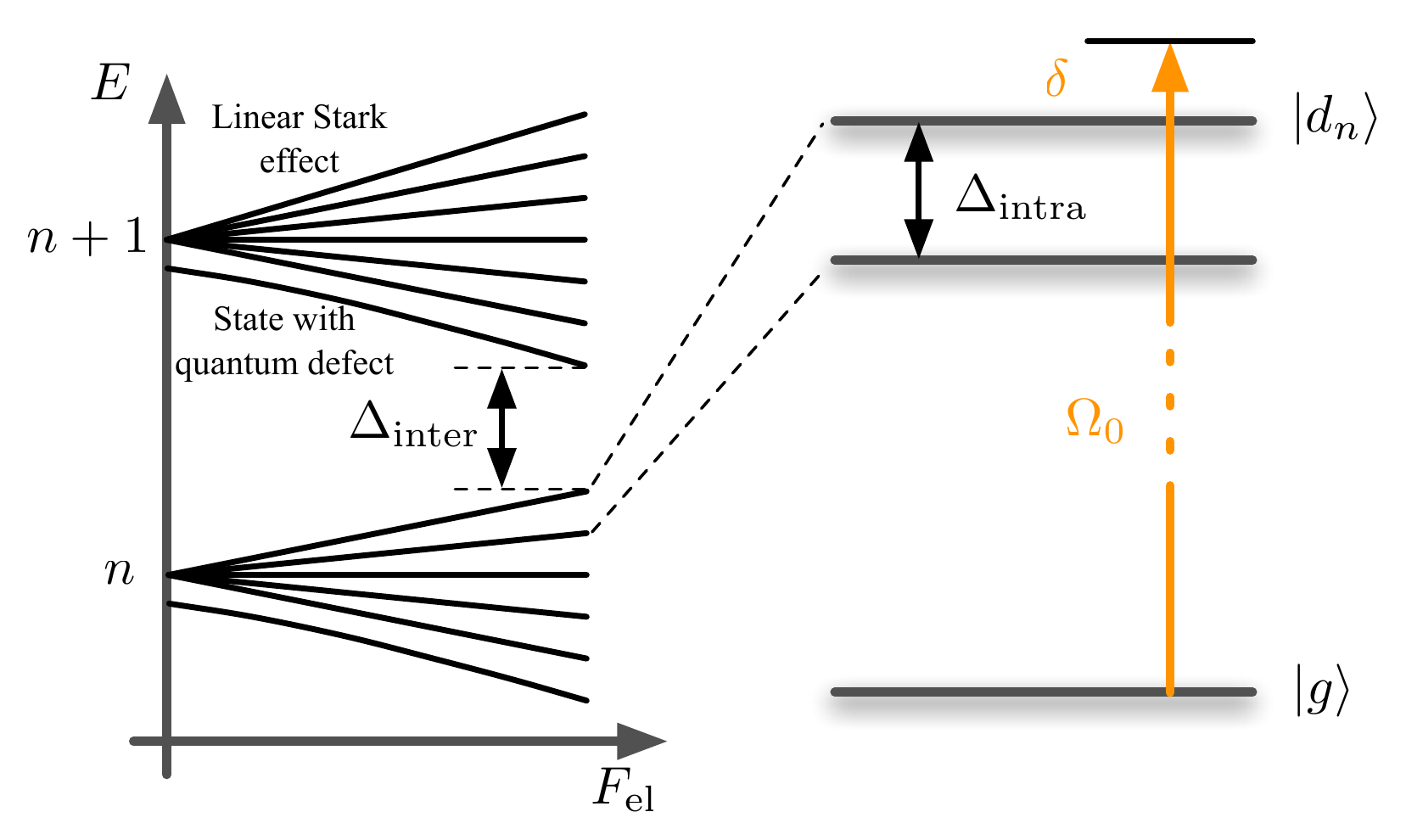}
\caption{Sketch of the level structure of an alkali metal atom as a
  function of the electric field strength. The energy of hydrogen-like
  states grows linearly with the field strength $F_\mathrm{el}$. Low
  angular momentum states with quantum defect exhibit a second order
  Stark effect. In our scheme we couple the ground state
  $\left|g\right>$ of an atom to the Rydberg state $\left|d_n\right>$,
  which exhibits the strongest Stark shift within the given
  $n$-manifold. The magnetic field strength and the laser parameters
  are set such that $\Delta_\mathrm{intra}\lesssim
  \Delta_\mathrm{inter}$, $\Delta_\mathrm{intra}\gg |\delta|$ and
  $\Delta_\mathrm{intra}\gg \Omega_0$.}\label{fig:level_structure}
\end{figure}

A sketch of a Rydberg-Stark level structure is shown in
figure~\ref{fig:level_structure}. For sufficiently small field
strengths, i.e. below the Inglis-Teller limit where the atom becomes
ionized, the levels are grouped in manifolds which can be labelled by
the principal quantum number $n$.  In our study we focus on the
energetically highest state of such a manifold whose azimuthal quantum
number is $m=0$. Employing the formula for hydrogen, the electric
dipole moment of this state, which we denote as $\left|d_n\right>$,
evaluates to
\begin{align}
  \mathbf{d}_n=d_n\mathbf{e}_z=\frac{3}{2}e a_0n(n-1)\mathbf{e}_z, 
  \label{eq:dipole_moment}
\end{align}
with $e$ and $a_0$ being the charge of the electron and the Bohr
radius, respectively.

Two atoms excited to the state $\left|d_n\right>_k$ and $\ket{d_n}_l$
at distinct sites of the lattice $k$ and $l$, are expected to interact
strongly via the dipole-dipole interaction potential, as already for
$n=17$ equation~(\ref{eq:dipole_moment}) gives a dipole moment of more
than one kiloDebye. We intend to work in a regime in which the
interaction can be treated within first order perturbation carried out
in single particle product states. To this end it is important to note
that the dipole-dipole interaction commutes with $L_{z_1}+L_{z_2}$,
and therefore only couples atomic pair states for which the azimuthal
quantum numbers obey $m_1+m_2=m_1^\prime+m_2^\prime$
\cite{Jaksch00}. We assume that all of these couplings are far
off-resonant, which can be assured by properly tuning
$\Delta_\mathrm{intra}$ and $\Delta_\mathrm{inter}$ using the electric
field. Moreover, accidental resonances can be avoided by restricting
oneself to low principal quantum numbers $n$ since here the
two-particle energy spectrum possesses large gaps which make
near-resonant excitations unlikely. Employing $\mathbf{d}_n \perp
\mathbf{e}_{x}$ the interaction Hamiltonian between two atoms being in
the product state $\left|d_n\right>_k\left|d_n\right>_l$ can then be
written as
\begin{align}
  \hat H^{\mathrm{int}}_{kl}
  &=
  2V_\mathrm{dd}^{(1)}\frac{d_n^2}{|k-l|^3}
  \ket{d_n}_k\ket{d_n}_l
  \bra{d_n}_k\bra{d_n}_l 
,
\end{align}
with $V_\mathrm{dd}^{(1)}\approx 1/(8\pi\epsilon_0a^3)$, when the confinement length in each lattice site is much smaller than the lattice spacing. 

We now turn to the description of the interaction between the laser
and the atom. We employ a two-level approximation assuming that the
ground state $\left|g\right>_k$ is coupled only to the state
$\left|d_n\right>_k$.  This is justified as long as the modulus of the
laser detuning $\delta$ and the Rabi frequency $\Omega_0$ are much
smaller than the energy gaps $\Delta_\mathrm{intra}$ and
$\Delta_\mathrm{inter}$.  Making the rotating wave approximation, the
interaction of the laser with an atom located on the $k$-th site can
be written as ($\hbar=1$)
\begin{align}
  \hat H^{\mathrm{L}}_{k}=
  \Omega_0
 \left(
\ket{d_n}_k\bra{g}_k +
\ket{g}_k\bra{d_n}_k 
\right)
  -
  {\delta}
  \ket{d_n}_k\bra{d_n}_k
\end{align}
with $\Omega_0$ and $\delta$ being the Rabi frequency and the
detuning, respectively.

\subsubsection{Many-body model}

We now consider the many-body physics when many atoms in an optical
lattice are simultaneously coupled to the state
$\left|d_n\right>_k$. We assume that either atoms are loaded into a
optical lattice with tight radial confinement, so that we have one
atom per site along a 1D tube, or that we have weak radial confinement
and many atoms per site. In each case, we can assume that the
ground-state atoms are confined to a single modefunction at each given
lattice site, $\phi_k(\mathbf{x})$. In the case of a single atom per
site, this is the Wannier function for the lowest Bloch band, and in
the case of many (weakly-interacting) atoms it is the solution to the
Gross-Pitaevskii equation in the mean-field limit.

In order to formulate the Hamiltonian of the many-body system, we
introduce field operators $\hat \Psi_g(\mathbf{x})$ and $\hat
\Psi_r(\mathbf{x})$, corresponding respectively to atoms in the ground
and Rydberg states, and obeying the standard bosonic (or fermionic) commutation relations. 
We assume that the timescale for excitation to the
Rydberg state is much shorter than timescales corresponding to the
atomic motion, and that therefore we can neglect population of any
modefunction other than the $\phi_k(\mathbf{x})$ for either the ground
or Rydberg states.  In this case, we can approximately expand the
field operators in terms of mode operators corresponding to each of
these modes,
\begin{align}
  \hat \Psi_g(\mathbf{x})&=\sum_k \phi_k(\mathbf{x}) \hat g_k\\
  \hat \Psi_r(\mathbf{x})&=\sum_k \phi_k(\mathbf{x}) \hat r_k,
\end{align}
where both the initial field operators and the mode operators, $\hat
g_k^\dag$ and $\hat r_k^\dag$, obey the appropriate bosonic
commutation or fermionic anti-commutation relations, depending on which
atoms are used in the experiment. We can then write the Hamiltonian
for a lattice with $N$ sites as
\begin{align} \label{eq:pre_working_hamiltonian}
  \hat H_{N}&\approx
\Omega_0 \sum^N_k
 \left[\hat g_k \hat r^\dagger_k+
    \hat g^\dagger_k \hat r_k\right]
-
  \delta\sum^N_k \hat n_k 
 \nonumber\\
  &\qquad+\sum^N_k V^\mathrm{on}_{\mathrm{dd}}\hat n_k( \hat n_k-1) 
  + \frac{V_{\mathrm{dd}}}{2}\sum^N_{k\neq l} \frac{\hat n_k \hat n_m}{|k-l|^3}
\end{align}
with $\hat n_k= \hat r^\dagger_k \hat r_k$, $V_{\mathrm{dd}}\approx
d_n^2/4\pi \epsilon_0 a^3$. In the last term we have again neglected
corrections to the shape of the interaction potential, which are of
the order $\order{l_\mathrm{H}/(a|k-l|)}$, with $l_\mathrm{H}$ the
characteristic harmonic oscillator length for the deep optical
potential well on each site.

The quantity $V^\mathrm{on}_{\mathrm{dd}}$ is relevant only in the
case that we initially have many atoms per site, and gives the energy
offset due to the dipole-dipole interaction if two Rydberg atoms are
excited on the same site. We will assume that
$V^\mathrm{on}_{\mathrm{dd}} \gg V_{\mathrm{dd}}/|k-l|^3$ for $k\neq
l$ and $\Omega_0, (N\delta) \ll V^\mathrm{on}_{\mathrm{dd}}$, so that
the occupation of a single site by two Rydberg atoms is far
off-resonant and therefore the excitation of such a state is highly
improbable. This phenomenon, which is known as the Rydberg blockade,
makes it possible to restrict the maximum number of Rydberg atoms per
site to one, such that $\hat n_k$ can only assume the two eigenvalues
$0$ and $1$.

Secondly, to arrive at a Hamiltonian governing solely the dynamics of
the Rydberg excitations, we eliminate the creation and annihilation
operators $\hat g_k^\dag$ and $\hat g_k$ in
equation~(\ref{eq:pre_working_hamiltonian}). In the case where we have
an initial ground state with a single atom per site, this is simply a
relabelling of the state with a single ground state atom per site as
the vacuum state of the operators $r_k$. In the case where the mode
function at each site is occupied by the same macroscopic number
$N_\mathrm{g}\gg 1$, on the other hand, we make the replacement $\hat
g_k, \hat g_k^\dagger \rightarrow \sqrt{N_\mathrm{g}}$, in which case
we define the effective Rabi frequency as $\Omega = \Omega_0
\sqrt{N_\mathrm{g}}$. The model we derive is exact under the stated
assumptions when we have a fixed number of atoms per site (e.g.,
beginning deep in the Mott Insulator regime for bosons or in a band
insulator for Fermions), which is the ideal experimental
realization. An approximation to the model can also be obtained in a
weakly interacting gas, but with some inhomogeneity in $\Omega$ for
different lattice sites, both due to position-dependence of the mean
density (due to a trapping potential), and also due to number
fluctuations between sites. In production of crystalline states via
the adiabatic ramping process discussed below, we expect that the ramp
is somewhat robust against these fluctuations, especially for high
filling factor crystalline states. However, this depends on the
details of the realization and the inhomogeneity in $\Omega$.

Thus, we arrive at the spin Hamiltonian describing excitation of Rydberg states, 
\begin{align} \label{eq:rydhamiltonian}
  \hat H_\mathrm{ryd}=
  \Omega \sum^N_k \hat \sigma^x_k
  +\delta\sum^N_k \hat n_k
  +\frac{V_{\mathrm{dd}}}{2}\sum^N_{k\neq l} \frac{\hat n_k \hat n_l}{|k-l|^3},
\end{align}
where the Pauli matrix $\hat \sigma^x_k \equiv \hat r_k^\dag + \hat
r_k$ has been defined. This Hamiltonian is equivalent to our model
system from equation~(\ref{eq:hamiltonian}), with the states corresponding
to $\ket{\alpha}_k\equiv \ket{g}_k$ and $\ket{\beta}_k\equiv
\ket{d_n}_k$.

The scheme presented in section~\ref{sec:dynam-cryst-state} achieves a
crystalline phase of Rydberg excitations on very short
time\-scales. However, whilst the preparation can be very fast compared
to the lifetime of the Rydberg states, this lifetime will pose an
ultimate limit to the stability of the many-body state that is
produced. One way around this is to transfer the excited Rydberg
states to alternative stable ground states. However, an important
issue arises in this case because of the momentum kicks that arise due
to the interactions between excited atoms, which can create very
significant excitations in the motional states of the atoms. This is
especially an issue towards the end of the chain, where the atoms will
experience a force primarily in the direction away from the centre of
the chain. These kicks can be estimated classically from the force
acting due to the dipole-dipole interaction potential between excited
Rydberg state. For crystalline states one can estimate that the net
momentum kick during the crystal preparation time decreases as $k_R
\propto 1/ (f i)^4$, where $i$ denotes the $i$-th excited Rydberg atom
from the boundary and $f$ the periodicity of the crystal, i.e. with
$N/f$ filling. Quantitatively, transitions to higher bound states are
given by the size of matrix elements $\int w_n(x) w_0(x) e^{i k_R x}
dx$, where $w_n$ $(n>0)$ denotes the higher band Wannier
modefunctions. Thus the transition effects become small in the case
that $k_R a_0$, where $a_0$ denotes the oscillator length, is
small. The oscillator length is typically of the order of tens of nm
and we can estimate the momentum kick of the boundary excitation for a
$N/2$ filling crystal as roughly $1/$nm. Thus, this will be a problem
for atoms at the end of the chain in the $N/2$ case, but decreases
rapidly for atoms away from the edge, or a larger periodicity in the
crystal. One way to circumvent this problem for the case where all
lattice sites are singly occupied would be to use induced dipoles
rather than direct coupling to Rydberg levels from the outset, as
discussed in reference~\cite{Pupillo08}.  The idea is to make use of
two stable ground states, one of which is coupled off-resonantly to a
Rydberg level, producing an induced dipole with a strength depending
on the parameters of the coupling and the static dipole of the Rydberg
level. In this way, the coupling between ground and Rydberg states in
our model is replaced by the coupling between two stable ground
states. This can be achieved, e.g., via a two-photon transition.  This
results in much smaller values of $\Omega$ and $\delta$ (and hence
also the interaction strength between excited atoms), which could even
be made comparable to or less than the band gap in the optical
lattice. The atoms would then always remain trapped in the
lattice. The disadvantage of this method would be substantially
increased times for the total adiabatic process, however, the ratio of
the preparation time to spontaneous emission lifetime of the state
could always be made large.

This method would have the added benefit of a much smaller linewidth
of the coupling field for the transition between interacting and
non-interacting states. Especially for crystal periodicities other
than two lattice sites, the narrow regions of detuning $\delta/V_{\rm
  dd}$ for which the phases are found implies a strong condition on
the stability of the laser detuning in an experiment (and hence on the
linewidth of the exciting laser). For example, a crystalline state
with an excitation on every fourth site can be found in the region of
$\Omega/V_\mathrm{dd}=4\times10^{-3}$ for $0.06 < \delta/V_\mathrm{dd}
< 0.09$.  Using $V_{\mathrm{dd}}\approx 7 \hbar \, \mathrm{GHz}$ from
the previous example, this means that the range of detuning values in
which states of the character is found is approximately $200\,
\mbox{MHz}$, and we require that the laser linewidth is smaller than
this value.  This is realistic, however for crystalline phases with
even larger spacings between the excitations, the requirements become
even stronger.  If we use dressed dipoles, then we would be using a
two-photon transition to couple between the states, for which the
detuning can be controlled very accurately.

\section{Conclusion and Outlook}

\label{txt:outlook}

In conclusion, using adiabatic sweeps of microwave or laser
parameters, it is possible to form crystalline phases of strongly
interacting states in either polar molecules or neutral atoms coupled
to Rydberg states in an optical lattice. The effective model
Hamiltonian exhibits a rich ground state phase diagram, with the
crystalline order being destroyed as the coupling between states is
increased. For typical experimental size scales and parameters we
showed that preparation of crystalline structures can be achieved on
reasonable timescales, and the structures can be detected by
measurement of characteristic density-density correlation functions.

The results we have obtained here motivate further study of the phase
diagram with its lobe-like structure of different crystalline phases
(as indicated in figure~\ref{fig:gap}). This also opens the
opportunity to extend 1D time-dependent simulations to the study of
crystal formation in regimes of significantly larger spacing than is
studied here (e.g., using matrix product operator methods
\cite{Verstraete04, Zwolak04, Mcculloch07, Crosswhite08, Pirvu10}) and
also to probe time-dependent transitions between the different
phases. This will also allow connections to current work in which
Rydberg excitations of randomly distributed atoms are studied
\cite{Pohl10}.

\begin{acknowledgments}
  We thank Hans Peter B\"uchler, Ignacio Cirac, Markus M\"uller,
  Beatriz Olmos Sanchez, Thomas Pohl, Hendrik Weimer, and Peter Zoller
  for helpful discussions. We thank Alexey Gorshkov for helpful
  discussions, and for pointing out a small error in our original
  parameter estimation for polar molecules.  This work was supported
  by the Austrian Science Fund (FWF) through the SFB FOQUS and the
  I118\_N16 (EuroQUAM\_DQS) project, by the DARPA OLE program, the
  Austrian Ministry of Science BMWF as part of the
  UniInfrastrukturprogramm of the Forschungsplattform Scientific
  Computing and of the Centre for Quantum Physics at LFU Innsbruck,
  and AFOSR MURI, EOARD grant number FA8655-10-1-308.
\end{acknowledgments}

\end{document}